\documentclass[aps,prd,twocolumn,reprint,superscriptaddress,nofootinbib,floatfix]{revtex4-2}
\RequirePackage[l2tabu, orthodox]{nag}

\usepackage[utf8x]{inputenc}
\usepackage{microtype} 

\usepackage{amsmath}
\usepackage{amstext, amssymb, amsthm, amsfonts, slashed}
\usepackage{physics}
\usepackage{mathtools}
\usepackage{mhchem}
\usepackage{graphicx}
\usepackage{booktabs}
\usepackage{dcolumn}% Align table columns on decimal point
\usepackage{bm}% bold math
\usepackage{dsfont}
\usepackage[usenames,dvipsnames]{xcolor}
\usepackage[normalem]{ulem}
\usepackage{footmisc}

\usepackage{url}
\usepackage[colorlinks=true,urlcolor=blue,linkcolor=blue,citecolor=blue,hypertexnames]{hyperref}
\usepackage[nameinlink]{cleveref}
\usepackage{braket}
\usepackage{simplewick}
\usepackage{float}
\usepackage{multirow}

\hypersetup{
	colorlinks,
	linkcolor={blue!75!black},
	citecolor={blue!75!black},
	urlcolor={blue!75!black}
}
\usepackage{pifont}% http://ctan.org/pkg/pifont

\def \psib{\bar{\psi}}

\crefname{section}{Sec.\!}{Secs.\!}
\crefname{figure}{Fig.\!}{Figs.\!}
\crefname{equation}{}{}
\crefname{table}{Tab.\!}{Tabs.\!}
\crefname{appendix}{App.\!}{Apps.\!}

\newcommand{\GNewton}{\ensuremath{G_{\mathrm{N}}}}
\newcommand{\CC}{\ensuremath{\Lambda}}

\newcommand{\gnewton}{\ensuremath{g}}

\newcommand{\sigmaFRic}{\ensuremath{\sigma_{\mathrm{Ric}}}}
\newcommand{\sigmaFR}{\ensuremath{\sigma_{\mathrm{R}}}}

\newcommand{\yukR}{\ensuremath{y_{\mathrm{R}}}}
\newcommand{\yukRsq}{\ensuremath{y_{\mathrm{R^2}}}}
\newcommand{\yukCsq}{\ensuremath{y_{\mathrm{C^2}}}}

\newcommand{\orcid}[1]{\href{https://orcid.org/#1}{\includegraphics[height=1.7ex,width=1.7ex]{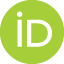}}}

\begin{document}

%%%%%%%%%%%%%%%%%%%%%%%%%%%%%%%%%%%%%%%%%%%
% opening
%%%%%%%%%%%%%%%%%%%%%%%%%%%%%%%%%%%%%%%%%%%
\title{Yukawa interactions in Quantum Gravity}

\author{Gustavo~P.~de~Brito~\orcid{0000-0003-2240-528X}}
\thanks{\href{mailto:gp.brito@unesp.br}{gp.brito@unesp.br}} 
\affiliation{Departamento de F\'isica, Universidade Estadual Paulista (Unesp), Campus Guaratinguet\'a,
	Av.~Dr.~Ariberto Pereira da Cunha, 333, Guaratinguet\'a, SP, Brazil.}
\author{Manuel~Reichert~\orcid{0000-0003-0736-5726}}
\thanks{\href{mailto:m.reichert@sussex.ac.uk}{m.reichert@sussex.ac.uk}} 
\affiliation{Department  of  Physics  and  Astronomy,  University  of  Sussex,  Brighton,  BN1  9QH,  U.K.}
\author{Marc Schiffer~\orcid{0000-0002-0778-4800}}
\thanks{\href{mailto:marc.schiffer@ru.nl}{marc.schiffer@ru.nl}} 
\affiliation{Institute for Mathematics, Astrophysics and Particle Physics, Radboud University,
	Heyendaalseweg 135, 6525 AJ Nijmegen, The Netherlands.}

\begin{abstract}
We present the first complete next-to-leading-order analysis of a Yukawa system within the framework of asymptotically safe quantum gravity. Our results are obtained through a systematic resummation of higher-order operators, revealing two distinct resummation mechanisms -- one of which has not been explored previously. In addition, we introduce a novel approach to estimate systematic uncertainties by simulating the impact of neglected higher-order contributions.

We demonstrate that quantum gravity fluctuations anti-screen Yukawa interactions, thereby resolving previously inconclusive leading-order results. This anti-screening mechanism enables the generation of finite interactions from an asymptotically free Yukawa fixed point. Consequently, our findings provide strong evidence that non-vanishing Yukawa couplings are compatible with asymptotically safe quantum gravity, which is a necessary requirement for the Standard Model to emerge from an asymptotically safe ultraviolet completion.
\end{abstract}

\maketitle

%%%%%%%%%%%%%%%%%%%%%%%%%%%%
\textit{Introduction.---} 
%%%%%%%%%%%%%%%%%%%%%%%%%%%%%%%%%
Deriving testable predictions or implications from a quantum theory of gravity is a big challenge. The absence of direct experimental access to the quantum-gravitational regime requires us to resort to consistency tests, such as formal, internal criteria to ensure the theory is unitary, background independent, etc.. Importantly, the theory should describe nature and be consistent with experimental observations at low energies. The matter sector offers a large test-ground for any fundamental theory: whatever the properties in the deep ultraviolet (UV), the matter content of the Standard Model (SM) with possible extensions, together with all interactions and masses, has to emerge at low energies.

In this letter, we focus on the emergence of Yukawa interaction in asymptotically safe quantum gravity \cite{Weinberg:1980gg, Reuter:1996cp}. Asymptotic safety is built upon quantum scale invariance at high energies, characterised by an interacting renormalisation group fixed point in the UV, see \cite{Bonanno:2020bil, Dupuis:2020fhh, Knorr:2022dsx, Eichhorn:2022gku, Morris:2022btf, Wetterich:2022ncl, Martini:2022sll, Saueressig:2023irs, Pawlowski:2023gym, Platania:2023srt, Bonanno:2024xne, Reichert:2020mja, Basile:2024oms} for reviews and lecture notes. By now, there is compelling evidence that the fixed point exists for SM matter content and beyond \cite{Dona:2013qba, Meibohm:2015twa, Biemans:2017zca, Christiansen:2017cxa, Alkofer:2018fxj,  Eichhorn:2018akn, Eichhorn:2018ydy, Eichhorn:2018nda, Burger:2019upn, Wetterich:2019zdo, deBrito:2020xhy, Pastor-Gutierrez:2022nki, Korver:2024sam}. Furthermore, qualitative mechanisms have been found that provide a UV completion for all SM couplings. They rely on a gravitational anti-screening effect on the Yukawa and Abelian gauge coupling \cite{Eichhorn:2017ylw, Alkofer:2020vtb, Pastor-Gutierrez:2022nki, Eichhorn:2025sux}, or, in the case of the Yukawa coupling, a small gravitational screening compensated by a larger anti-screening from the gauge coupling \cite{Eichhorn:2018whv, Eichhorn:2025sux}. Combining this with a screening gravitational contribution to the quartic scalar coupling allows for a prediction of the Higgs mass \cite{Shaposhnikov:2009pv}. Indeed, computations indicate that the gravitational contribution to the Abelian gauge coupling is anti-screening \cite{Daum:2009dn, Harst:2011zx, Folkerts:2011jz, Christiansen:2017gtg, Eichhorn:2017ylw, Eichhorn:2017lry, Christiansen:2017cxa, Eichhorn:2019yzm, deBrito:2019umw, Pastor-Gutierrez:2022nki, deBrito:2022vbr, Riabokon:2025ozw}, and that the gravitational contribution to the quartic coupling is screening \cite{Narain:2009fy, Percacci:2015wwa, Labus:2015ska, Oda:2015sma, Hamada:2017rvn, Eichhorn:2017als,  Eichhorn:2017ylw, Pawlowski:2018ixd,Wetterich:2019rsn, Eichhorn:2020sbo, deBrito:2019umw, deBrito:2022vbr}.

For the Yukawa couplings, the status is unclear on a qualitative level as the gravitational contribution can be either screening or anti-screening, depending on the gravitational fixed-point values \cite{Zanusso:2009bs, Oda:2015sma, Eichhorn:2016esv, Hamada:2017rvn,  Eichhorn:2017eht, deBrito:2019umw, Pastor-Gutierrez:2022nki, deBrito:2022vbr}. A strong screening contribution would imply a no-go theorem for fundamental Yukawa couplings in asymptotic safety. The inconclusive status warrants a comprehensive study of the robustness of current results.

In this work, we provide the first analysis of the Yukawa sector, including the complete set of operators that contribute to the Yukawa coupling at next-to-leading order (NLO) in the Newton coupling. We discover a novel mechanism on how asymptotically free higher-order operators give a resummed contribution to the critical exponent of the Yukawa coupling, and provide a new method to estimate systematic uncertainties. Our results show that metric fluctuations anti-screen the Yukawa interaction. We therefore provide strong evidence that non-vanishing Yukawa interactions can emerge from an asymptotically safe UV completion.

%%%%%%%
\begin{figure*}[tbp]
\includegraphics[width=.9\linewidth]{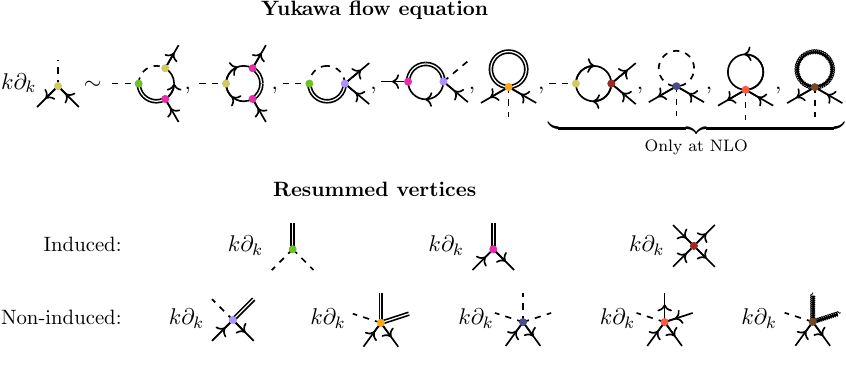}
\caption{We show the diagrammatic flow of the Yukawa coupling. At LO, all vertices and propagators are given by their tree-level expressions. For NLO, we include the flow equations of all vertices and propagators that appear in the LO flow of the Yukawa coupling and the scalar and fermion propagators.}
\label{fig:Yukawa-flow}
\end{figure*}
%%%%%%%%

\bigskip
%%%%%%%%%%%%%%%%%%%%%%%%%%%%
\textit{Setup.---} 
%%%%%%%%%%%%%%%%%%%%%%%%%%%%%%%%%
We consider a Standard-Model-like theory with Yukawa interactions under the influence of quantum gravity. The action is composed of the Einstein-Hilbert action and a matter action, $S = S_\text{EH} + S_\text{matter}$, with
\begin{align} \label{eq:EH-action}
S_\text{EH}=\frac{1}{16\pi\,\GNewton}  \int \mathrm d^4 x \sqrt{g}\left(2\CC - R\right).
\end{align}
The matter part is given by
\begin{align} \label{eq:matter-action}
S_\text{matter} &= \int \mathrm d^4 x \sqrt{g}\,\bigg(\frac{1}{2}\,\partial_{\mu}\phi_I\,\partial^{\mu}\phi_I+\frac{1}{4}F^{a \mu\nu}F^{a}_{\mu\nu} \notag \\
&\qquad \qquad \qquad +i\bar{\psi}_i \slashed{D}\psi_i + iy \,\phi_1 \psib_i\psi_i\bigg),
\end{align}
where the index $I$ in $\phi_I$ labels $N_{\mathrm{s}}$ scalar fields,  the index $i$ in $\psi_i$ labels $N_{\mathrm{f}}$ fermionic fields, and the index $a$ in $F_{\mu\nu}^{a}$ labels $N_{\mathrm{v}}$ gauge fields. Here we focus on minimal matter (MM) ($N_\text{s}=1$, $N_\text{f}=1$, and $N_\text{v}=0$) as well  SM matter content ($N_\text{s}=4$, $N_\text{f}=22.5$, and $N_\text{v}=12$). In \cref{eq:matter-action}, we have introduced the Yukawa coupling with respect to one scalar field $\phi_1$, while there is a flavour symmetry in the fermion sector.

In this work, we determine the gravitational contribution to the Yukawa coupling. At leading order in $G_\text{N}$, this was previously done in perturbation theory \cite{Rodigast:2009zj}, and with the functional renormalisation group (FRG) \cite{Zanusso:2009bs, Oda:2015sma, Eichhorn:2016esv, Hamada:2017rvn,  Eichhorn:2017eht, deBrito:2019umw, Pastor-Gutierrez:2022nki, deBrito:2022vbr}, which can also include higher-order effects. Here, we extend the analysis to include the complete set of NLO operators as well as the induced higher-order resummations utilising the FRG.

The FRG is based on a modified dispersion $p^2 \longrightarrow p^2 + R_k(p^2)$ where the regulator $R_k$ acts as an infrared (IR) cutoff and implements the Wilsonian integrating out of momentum shells. The central object of the FRG is the scale-dependent effective action $\Gamma_k$, which includes all quantum fluctuations above the momentum scale $k$. The Wetterich equation describes how $\Gamma_k$ changes upon integrating out further modes \cite{Wetterich:1992yh, Morris:1993qb, Ellwanger:1993mw},
\begin{align}
k\partial_k \Gamma_k=\frac{1}{2} \mathrm{STr}\left[\left(\Gamma_k^{(2)}+\mathcal{R}_k\right)^{-1} k\partial_k \mathcal{R}_k\right].
\end{align}
Here $\Gamma_k^{(2)}$ is the second functional derivative of $\Gamma_k$ with respect to all fields, and $\mathcal{R}_k$ is the regulator matrix with entries $R_k$. For a review on the FRG, see \cite{Dupuis:2020fhh}. In asymptotic safety, the dimensionless versions of all couplings realise a fixed point in the UV. Therefore, we introduce the dimensionless counterparts of the Newton coupling and cosmological constant as 
\begin{align} \label{eq:dimless-couplings}
g &= G_\text{N} k^2\,,
&
\lambda &= \Lambda/k^2\,,
\end{align}
respectively.

\bigskip
%%%%%%%%%%%%%%%%%%%%%%%%%%%%
\textit{Leading order quantum gravity contribution to the Yukawa coupling.---} 
%%%%%%%%%%%%%%%%%%%%%%%%%%%%%%%%%
With the action given by \cref{eq:EH-action,eq:matter-action}, the beta function of the Yukawa coupling can be schematically written as
\begin{align}
\beta_{y,\,\mathrm{LO}} \equiv k \frac{\partial y}{\partial k} = c_y y^3 - f_y\, \gnewton \, y\,,
\label{eq:beta-yuk-schematic}
\end{align}
where the coefficient $f_y$ parameterises the gravitational contribution \cite{Zanusso:2009bs, Oda:2015sma, Eichhorn:2016esv, Eichhorn:2017eht, deBrito:2022vbr, Pastor-Gutierrez:2022nki}. At leading order, the gravitational contribution is linear in $\gnewton$, and $f_y$ explicitly depends on $\lambda$.\footnote{Note that the Newton coupling is often absorbed into $f_y$ in the literature.}  However, $f_y$ can encode higher orders as it can depend on further powers of $g$.

The key question is whether gravitational fluctuations anti-screen the Yukawa interaction. This is realised if the fixed point $y^*=0$ is relevant in asymptotic safety. In this case, the Yukawa coupling is asymptotically free and UV complete, but finite Yukawa interactions are generated towards the IR. Whether a coupling is relevant is determined by the critical exponent, which is given at leading order by
\begin{align}
\label{eq:thetaY-LO}
\theta_{y,\,\text{LO}} = - \partial_y \beta_y|_* = f_y \, g^* \,,
\end{align}
for the non-interacting fixed point $y_*=0$. We require $\theta_y>0$ for a relevant direction, which at leading order corresponds to $f_y>0$. Note that we have assumed that the Abelian gauge coupling becomes asymptotically free. If the gauge coupling instead becomes asymptotically safe, the critical exponent receives an additional contribution, and $\theta_y$ can still be positive even with a slightly negative gravitational contribution \cite{Eichhorn:2018whv, deBrito:2022vbr, Eichhorn:2025sux}. For our work, it is not necessary to distinguish these scenarios since the allowed negative range is tiny.

%%%%%%
\begin{table}[t]
\centering
\renewcommand{\arraystretch}{1.2}
\begin{tabular}{|c|c|c|c|c|}
	\hline
	Vertex &Operator &Coupling& Dim. &   Ref.   \\ \hline
	\multirow{2}{*}{$h_{\mu\nu} \bar\psi\psi$}  &$R^{\mu\nu}\psib \gamma_\mu D_\nu \psi$ & \sigmaFRic &6 &   \cite{Eichhorn:2018nda} \\ 
	\cline{2-5} 
	&$R\,\psib  \slashed{D} \psi$ & \sigmaFR &6    & -   \\ \hline 
	\multirow{2}{*}{$h_{\mu\nu} \phi^2$} &$R^{\mu\nu}\, \partial_\mu\phi  \partial_\nu\phi$ & $\rho_\text{Ric}$&6    &  \cite{Eichhorn:2017sok, Laporte:2021kyp} \\ 
	\cline{2-5}  
	&$R\, \partial_\mu\phi  \partial^\mu\phi$ & $\rho_\text{R}$ &6    & \cite{ Laporte:2021kyp} \\ \hline 
	$(\psib \psi)^2$&$(\psib \gamma_\mu\psi)^2+(i\,\psib \gamma_\mu\gamma_5\psi)^2$ & $\lambda_{+}$ &6    & \cite{Eichhorn:2011pc, Meibohm:2016mkp}  \\ \hline\hline
	$\phi^2 \psib \psi$&$ \partial_\mu\phi  \partial^\mu\phi \, \psib  \slashed{D} \psi$ & $\chi_i$ &8    & \cite{Eichhorn:2016esv}  \\ \hline
	$\phi^4$&$(\partial_\mu\phi  \partial^\mu\phi)^2$ & $\omega_i$ &8    & \cite{Eichhorn:2012va, Laporte:2021kyp, Knorr:2022ilz, deBrito:2023myf}   \\ \hline
	$h_{\mu\nu}^2 \bar\psi\psi$ & $R^2\,\psib  \slashed{D} \psi$  &$\sigma_{R^2,i}$ & 8 & - \\ \hline
	$h_{\mu\nu}^2 \phi^2$ & $R^2 \partial_\mu\phi  \partial^\mu\phi$  & $\rho_{R^2,i}$  & 8 & - \\ \hline
\end{tabular}
\caption{All operators that contribute explicitly to $\beta_y$ (upper half) and to the flow of the scalar and fermion anomalous dimension (lower half). These operators do not possess a Gau\ss ian fixed point and hence they are asymptotically safe. For the lower half, we only show one operator per vertex, as they are sub-leading.}
\label{tab:induced}
\end{table}
%%%%%%

Beyond leading order, the critical exponents are given by
\begin{align} \label{eq:def-stab-matrix}
\theta_i&=-\mathrm{Eig}\left(M_{jm}\right),
&
M_{jm}&=\frac{\partial \beta_j}{\partial g_m}\bigg|_{*}\,,
\end{align}
where we have introduced the stability matrix $M_{ij}$, which consists of the derivatives of all beta-functions with respect to all couplings.

Computations within the FRG framework indicate that $\Theta_{y,\, \mathrm{LO}}<0$. In our scheme, the values for minimal matter and SM read
\begin{align}
\label{eq:thetaY_LO}
\Theta_{y,\text{SM},\text{LO}} &= -0.22\,,
&
\Theta_{y,\text{MM},\text{LO}} &= -0.58\,.
\end{align}
It should be noted that the sign depends on the fixed-point value of the cosmological constant $\lambda$. In particular, $\lambda^*\lesssim-3$ is necessary to achieve $\theta_y>0$. This is contradicted by the fixed point value of the cosmological constant, which is located at positive values $\lambda^* >0$. In some approximations, the fixed-point value for SM matter shifts towards $\lambda^*<0$ \cite{Dona:2013qba, Biemans:2017zca, Alkofer:2018fxj}, but typically the anomalous dimensions are large, indicating a breakdown of the approximation. At face value, the leading order results \cref{eq:thetaY_LO} predict vanishing Yukawa couplings at low energies, which is incompatible with observations. However, the leading-order result is sensitive to the inclusion of higher-order curvature operators \cite{Hamada:2017rvn, Eichhorn:2017eht, deBrito:2019umw, deBrito:2019epw} and to changes of the regulator \cite{Pastor-Gutierrez:2022nki}, motivating a study beyond leading order.

%%%%%%
\begin{table}[t]
\centering
\renewcommand{\arraystretch}{1.2}
\begin{tabular}{|c|c|c|c|c|}
	\hline
	Vertex & Operator & Coupling &  Dim. & Ref.  \\ \hline
	$\phi \bar\psi\psi$ & $\phi ( \psib \Box \psi + \Box \psib  \psi)$& $y_\Box$ &6  & -  \\ \hline
	$h_{\mu\nu} \phi \bar\psi\psi$& $R \,\phi \psib \psi$& \yukR & 6  & - \\ \hline
	\multirow{2}{*}{$h_{\mu\nu}^2 \phi \bar\psi\psi$}& $R^2 \,\phi \psib \psi$& \yukRsq &8  & - \\
	\cline{2-5} 
	& $C_{\mu\nu\rho\sigma}^2 \,\phi \psib \psi$& \yukCsq&8  & -  \\ \hline 
	$\phi^3 \psib \psi$ & $(\partial_\mu \phi)^2 \,\phi \psib \psi$ & $y_{\phi^2}$ &8  & -  \\ \hline 
	$\phi (\psib \psi)^2$ & $(\psib  \slashed{D} \psi) \,\phi \psib \psi$ & $y_{\psib \psi}$ &8  & -  \\ \hline 
	$A^2 \phi \psib \psi$ & $(F_{\mu\nu} F^{\mu\nu}) \,\phi \psib \psi$ & $y_{A^2}$ &8  & -  \\ \hline 
\end{tabular}
\caption{All operators that contribute explicitly to $\beta_y$ but are not induced by quantum gravitational fluctuations. These operators are asymptotically free, and they contribute to $\Theta_y$ via the resummation of the stability matrix.}
\label{tab:non-induced}
\end{table}
%%%%%%

\bigskip
%%%%%%%%%%%%%%%%%%%%%%%%%%%%
\textit{Resummation of higher-order operators.---} 
%%%%%%%%%%%%%%%%%%%%%%%%%%%%%%%%%
Within standard perturbation theory, the NLO contribution to $\Theta_y$ can be obtained by a two-loop computation of $\beta_y$ in combination with the two-loop wave function renormalisations of the fields. In contrast, the FRG provides a non-perturbative one-loop equation and beyond leading order results are obtained through the resummation of higher-order operators. We determine the full set of operators that contribute at NLO to $\Theta_y$ by analysing the flow equations for the Yukawa coupling, see \cref{fig:Yukawa-flow}. NLO operators are those that directly enter the vertices or propagators given in \cref{fig:Yukawa-flow}. We provide a comprehensive list of all NLO operators in \cref{tab:induced,tab:non-induced} and indicate which vertices they are entering. Note that more operators contribute to the resumming of vertices listed in \cref{tab:induced,tab:non-induced}, in particular related to different tensor contractions of already listed operators. We have checked explicitly that those operators do not contribute to $\Theta_y$ at NLO. Typically, the momentum or the Dirac structure causes these other operators to give no contribution. Furthermore, the coupling $y_{\phi^2}$ listed in \cref{tab:non-induced} does not contribute at NLO for our gauge-fixing choice, and therefore we neglect it in the following.

The operators that contribute at NLO can be divided into operators that are asymptotically free and safe. In \cref{tab:induced}, we list all operators that do not have an asymptotically free fixed point. Those operators share the symmetries of the kinetic terms of the gravity-matter system, and are therefore induced by gravitational fluctuations, see \cite{Eichhorn:2011pc, Eichhorn:2012va, Eichhorn:2016esv, Meibohm:2016mkp, Christiansen:2017gtg, Eichhorn:2017eht, Eichhorn:2017sok,  Eichhorn:2018nda, Eichhorn:2019yzm,  deBrito:2020dta, Eichhorn:2021qet, deBrito:2021pyi, Laporte:2021kyp, Knorr:2022ilz,  Eichhorn:2022gku, deBrito:2023myf}. The upper half includes all operators that directly enter the Yukawa diagrams, see \cref{fig:Yukawa-flow}, while the lower half lists all operators that only enter through the anomalous dimension.

The beta function of an induced operator with coupling $b$ has the following structure\footnote{For the four-point pure-matter vertices in \cref{tab:induced}, the structure is slightly different since the fixed point values are proportional to $g^2$ and the contribution to the Yukawa coupling is $g^0$. In the end, the structure of \cref{eq:cont-ind} also holds for these couplings.}
\begin{align}
\beta_b &= d_\mathcal{O} b + c_1 g  + \dots \,,
&
b^*& = - g^* \frac{c_1}{d_\mathcal{O} } + \mathcal{O}( {g}^2)\,,
\end{align}
where $d_\mathcal{O}$ is the mass dimension of the coupling $b$ associated with the operator $\mathcal{O}$, e.g., $d_\mathcal{O}=2$ for a dimension 6 operator. The term $c_1 g$ is responsible that the coupling $b$ does not have a free fixed point. Instead, it possesses a shifted Gau\ss ian fixed point \cite{Eichhorn:2011pc, Eichhorn:2012va, Eichhorn:2016esv, Meibohm:2016mkp, Christiansen:2017gtg, Eichhorn:2017eht, Eichhorn:2017sok,  Eichhorn:2018nda, Eichhorn:2019yzm,  deBrito:2020dta, Eichhorn:2021qet, deBrito:2021pyi, Laporte:2021kyp, Knorr:2022ilz,  Eichhorn:2022gku, deBrito:2023myf}.

The induced coupling $b$ contributes to $\beta_y$ with a term of the type $c_2\, g\, b\, y$. Therefore, in an expansion of Newton's coupling, the coupling gives a contribution to $\Theta_y$,
\begin{align} \label{eq:cont-ind}
\theta_y \big|_\text{ind} = - c_2\, g^*\,  b^* = - {g^*}^2 \frac{c_1\, c_2}{d_\mathcal{O}}  + \mathcal{O}( {g^*}^3)\,.
\end{align}
As expected, the contribution of these couplings starts at $\mathcal{O}(g^2)$ as it is a two-loop effect. Furthermore, the contribution is suppressed by the dimension of the coupling $d_\mathcal{O}$. Note that a dimension 8 operator is suppressed with an additional factor of 2 compared to a dimension 6 operator, but this can be compensated by the size of the coefficients $c_1$ and $c_2$.

Remarkably, also asymptotically free operators can influence the critical exponent of the Yukawa coupling. This is happening through a new mechanism that has not been explored yet in the literature. For such an effect, we need a coupling $\kappa$ for which 
\begin{itemize}
\item[(i)]$\kappa$ contributes linearly to $\beta_y$,
\item[(ii)]$y$ contributes linearly to $\beta_\kappa$.
\end{itemize}
In that case, the stability matrix linearised in $g$ takes the shape
\begin{align} \label{eq:stab-mat-schematic}
M= 
\begin{pmatrix}
	a_{11} \, \gnewton  & a_{12} \, \gnewton\\
	a_{21} \, \gnewton & d_{\mathcal{O}} + a_{22} \,\gnewton
\end{pmatrix},
\end{align}
where the off-diagonal elements indicate the mixing between the Yukawa and the asymptotically free coupling. At leading order, we have $a_{11}=-f_y$, and $a_{12}=a_{21}=0$ see \cref{eq:thetaY-LO}. Beyond leading-order, the diagonal coefficient is modified by the induced couplings, $a_{11}=-f_y+ c_2\,b+\dots$ see \cref{eq:cont-ind}, and $a_{12}\neq 0 \neq a_{21}$.

Computing the negative eigenvalues of \cref{eq:stab-mat-schematic} in an expansion of $g$, there is an additional contribution to the Yukawa critical exponent given by
\begin{align} \label{eq:cont-non-ind}
\theta_y \big|_\text{non-ind} = - {g^*}^2 \frac{a_{12}\, a_{21}}{d_\mathcal{O}} + \mathcal{O}( {g^*}^3) \,.
\end{align}
We see that the structure is identical to \cref{eq:cont-ind} despite the different origin. It is crucial that $a_{12} \neq 0 \neq a_{21}$, otherwise the stability matrix in \cref{eq:stab-mat-schematic} would be either upper or lower triangular and only the coefficient $a_{11}$ would contribute to $\Theta_y$. We list all operators that contribute via this mechanism in \cref{tab:non-induced}.

%%%%%%
\begin{table}[b]
\centering
\scalebox{.87}{
	\begin{tabular}{|c|c|c|c|c|c|c|c|c|c|c|c|}
		\hline
		&$y_R$ & $y_\Box$ & $y_{R^2}$ & $y_{C^2}$ & $y_{\bar\psi\psi}$ & $y_{A^2}$ &	$\sigma_\text{Ric}$ & $\sigma_\text{R}$ &$\rho_\text{Ric}$ & $\rho_\text{R}$ & $\lambda_+$ \\ \hline
		MM & $-4.1$ &  $ 0.35$ &$7.9$ &$-3.4$ & $0.74$ & $0$ & $-0.49$ & $0.33$ &  $-0.061$ & $0.079$ & $0.073 $ \\ \hline
		SM & $-13 $ & $0.81 $ & $32 $ &  $-17 $	& $40 $  & $1.8 $ & $-1.6$ & $1.0 $ & $-0.40$ & $0.53 $ & $0.13 $ \\ \hline
\end{tabular}}
\caption{Two-loop contribution of couplings to the Yukawa critical exponent. All numbers must be multiplied by $g^2$.}
\label{tab:loop-contribution}
\end{table}
%%%%%%

%%%%%%%
\begin{figure*}[tbp]
\includegraphics[width=\linewidth]{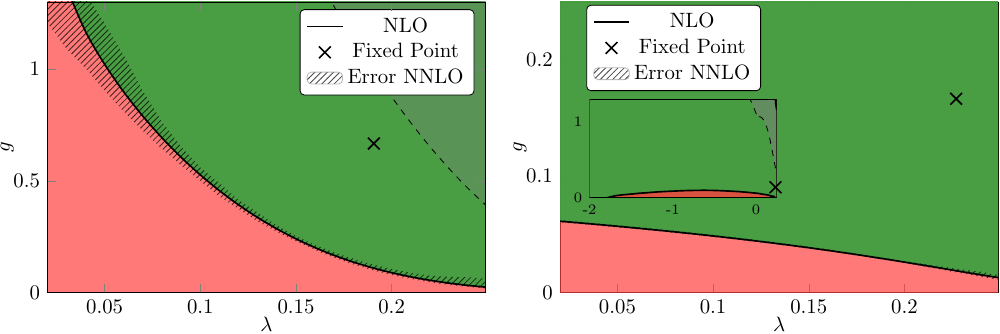}
\caption{We show the regions in the $\gnewton$-$\lambda$ plane where $\Theta_y>0$ (green) and $\Theta_y<0$ (red) together with the location of the UV fixed point for $N_\text{f}=N_\text{s}=1$ (left) and for SM matter content (right) at NLO. The hashed region is the estimated error by simulating the impact of an NNLO contribution. In the grey region above the dashed line, at least one anomalous dimension is large, $\eta >2$. The asymptotically safe fixed point is located in the green region with a relevant Yukawa coupling.}
\label{fig:G-Lambda-plot}
\end{figure*}
%%%%%%%

\bigskip
%%%%%%%%%%%%%%%%%%%%%%%%%%%%
\textit{UV fixed point.---} 
%%%%%%%%%%%%%%%%%%%%%%%%%%%%%%%%%
In the present work, we approximate the dynamics of the pure-gravity sector by the Einstein-Hilbert action, parameterised by the Newton coupling and cosmological constant. For the gravity-matter interplay, we include the upper half of \cref{tab:induced}, and all operators of  \cref{tab:non-induced}. The couplings in the lower half of \cref{tab:induced} only enter through the anomalous dimensions. Based on existing results \cite{Eichhorn:2016esv}, we have checked that these couplings are negligible. 

Within this setup, we find an interacting fixed point for gravity and all induced operators located at
\begin{align} \label{eq:FP_MM}
&(g,\, \lambda,\, \rho_\text{Ric},\, \rho_\text{R},\, \sigma_\text{Ric},\, \sigma _\text{R},\, \lambda_+)^*_\text{MM} \\
&\qquad=(0.67,\, 0.19,\, -0.12,\, 0.50,\, -0.030,\, 0.030,\, -1.0)\,, \notag 
\end{align}
for minimal matter, and at
\begin{align} \label{eq:FP_SM}
&(g,\, \lambda,\, \rho_\text{Ric},\, \rho_\text{R},\, \sigma_\text{Ric},\, \sigma _\text{R},\, \lambda_+)^*_\text{SM} \\	 
&\qquad	=(0.17,\, 0.23,\, -0.18,\, 0.37,\, -0.039,\, 0.039,\, -0.11)\,,\notag 
\end{align}
for SM matter. Conversely, all non-induced couplings are asymptotically free, $y_i^* \equiv 0$. 

Both fixed points feature three relevant directions, indicated by positive critical exponents, \cref{eq:def-stab-matrix}. Two are associated with the gravitational couplings. The third relevant direction is associated with the Yukawa coupling, and the critical exponents read
\begin{align} \label{eq:theta-Yuk-res}
\Theta_{y,\text{MM}} &= 3.1^{+1.8}_{-1.1} \,,
&
\Theta_{y,\text{SM}} &= 2.2^{+1.3}_{-1.0}  \,,
\end{align}
This is the main result of this work, and in contrast to the LO result, where the critical exponent was negative, see \cref{eq:thetaY_LO}. We have included an estimate of the error, and the computation thereof will be explained in a later section. The relevance of the Yukawa coupling implies that asymptotic safety resolves the Landau-pole problem of the Yukawa interaction by inducing asymptotic freedom. Consequently, finite Yukawa couplings, and hence finite fermion masses, at low energies can arise from a fundamental, asymptotically safe theory of quantum gravity and matter.

At the fixed points \cref{eq:FP_SM,eq:FP_MM}, all other critical exponents are close to the canonical mass dimension of the corresponding operators, with the values listed in the Supplemental Material. The existence of this fixed point, where higher-order operators feature near-canonical scaling, is non-trivial, and another important result of this letter. It further validates the existence of an asymptotically safe gravity-matter fixed point in an unprecedented approximation. The near-canonical scaling indicates the near-perturbative nature of the fixed point already observed in \cite{Falls:2013bv, Falls:2014tra, Denz:2016qks, Falls:2017lst, Falls:2018ylp, Eichhorn:2018akn, Eichhorn:2018ydy, Eichhorn:2018nda, Kluth:2020bdv}.

\bigskip
%%%%%%%%%%%%%%%%%%%%%%%%%%%%
\textit{Loop expansion.---} 
%%%%%%%%%%%%%%%%%%%%%%%%%%%%%%%%%
We expand $\Theta_y$, given in \cref{eq:theta-Yuk-res}, in $g$ at the fixed-point value of the cosmological constant $\lambda^*$. The contributions from each operator are listed in \cref{tab:loop-contribution}. For SM matter $y_{\psib \psi}$ provides the largest contribution, while for minimal matter $y_{R^2}$ is dominant. In both cases, the non-induced couplings provide a stronger contribution than the induced couplings, highlighting the importance of the off-diagonal contributions to the stability matrix.

Comparing both lines in \cref{tab:loop-contribution}, we observe that the absolute values of $y_R$, $y_{R^2}$, $y_{C^2}$, and $y_{\bar{\psi}\psi}$ grow significantly from the minimal matter to the SM content, while the other operators remain small. For the first three operators, this increase is caused by the shift of $\lambda^*$ to larger values, to which these operators are sensitive. For $y_{\bar{\psi}\psi}$, this increase is caused by an explicit contribution $\sim N_{f}$ in $\beta_y$, due to the closed fermion loop. Therefore, for large $N_f$, this operator gets enhanced compared to the curvature-Yukawa operators.

Summing up the contributions listed in \cref{tab:loop-contribution}, the loop expansion of $\Theta_y$ reads,
\begin{align} \label{eq:thetay_loop}
\Theta_{y,\text{MM}}&=  -0.88\,  g + 1.3\, g^2 + \mathcal{O}( {g}^3) \,, \notag \\
\Theta_{y,\text{SM}}&=  -1.3 \, g + 44\, g^2 + \mathcal{O}( {g}^3) \,.
\end{align}
These coefficients remain unchanged upon the inclusion of NNLO operators, apart from the changes induced by their dependence on $\lambda^*$. Remarkably, the NLO coefficient is positive for all values of the cosmological constant and for all matter content, as we have checked explicitly. Hence, the NLO contribution universally shifts the Yukawa coupling towards relevance.

Evaluating \cref{eq:thetay_loop} at the respective fixed point for $\gnewton$, see \cref{eq:FP_MM,eq:FP_SM}, shows that the strict quadratic expansion of $\Theta_y$ does not lead to $\Theta_{y,\text{MM}}>0$, instead $\Theta_{y,\text{MM}}\approx 0$.  Conversely, $\Theta_{y,\text{SM}}>0$ is already achieved in the strict quadratic expansion. This indicates that the SM system is more perturbative than the minimal matter system, as is also evident from the full set of critical exponents.

%%%%%%%
\begin{figure*}[tbp]
\includegraphics[width=\linewidth]{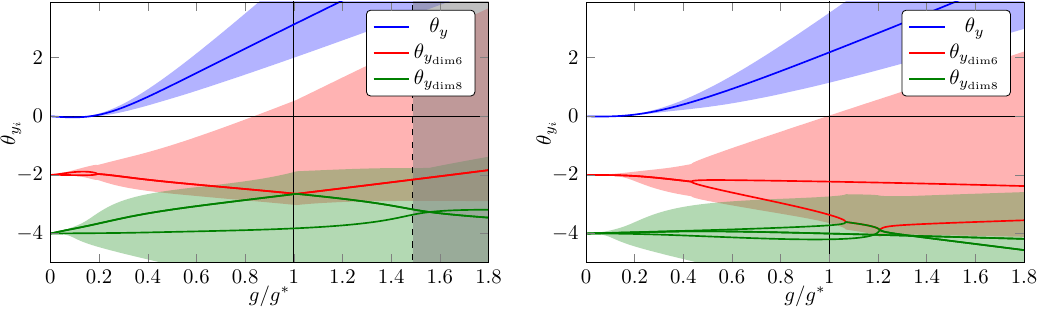}
\caption{We show the critical exponents of the Yukawa couplings with error estimates as a function of the Newton coupling at $\lambda=\lambda^*$ for $N_\text{f}=N_\text{s}=1$ (left) and for SM matter content (right).}
\label{fig:ThetasNLOvsNNLO}
\end{figure*}
%%%%%%%

\bigskip
%%%%%%%%%%%%%%%%%%%%%%%%%%%%
\textit{Error estimate.---} 
%%%%%%%%%%%%%%%%%%%%%%%%%%%%%%%%%
Next, we study the relevance of the Yukawa coupling as a function of the gravitational couplings to understand if the Yukawa coupling \textit{generically} becomes relevant at NLO, independent of the gravitational fixed-point values. Since the location of the fixed point is also subject to systematic uncertainties, this gives us some understanding of whether the qualitative result, namely that the Yukawa coupling becomes relevant in asymptotic safety, is robust under extensions of the approximation of the gravitational dynamics. In \cref{fig:G-Lambda-plot}, we show in green the parameter space for which the Yukawa coupling is relevant, while red indicates the parameter space for which the Yukawa coupling is irrelevant. The grey shaded regions indicate that an anomalous dimension becomes large, and hence the breakdown of the truncation. For minimal matter and SM, the fixed point is located well within the green region, such that moderate changes in $g^*$ and $\lambda^*$ do not change the relevance of the Yukawa coupling.

Besides the location of the fixed point, also the line separating the red and green regions in \cref{fig:G-Lambda-plot} is subject to systematic uncertainties: operators that contribute at NNLO via one of the two discussed mechanisms will change the location of that separation. As we have seen, the impact of non-induced operators is larger, so we focus on those operators. To estimate systematic uncertainties on $\Theta_y$ by neglecting operators such as $R^3\,\phi \psib \psi$, we simulate their effect on the level of the stability matrix, following the structure outlined in \cref{eq:stab-mat-schematic}: Any six-field vertex does not directly contribute to $\beta_y$, and hence the off diagonal term in the stability matrix vanishes, $a_{12}=0$, indicating the NNLO nature of those operators. Instead of computing the coefficients $a_{21}$ and the other off-diagonal elements of the full stability matrix, we estimate their effect via Monte-Carlo simulations: inspecting the full system at NLO, we determine the typical size of coefficients $a_{ij}$ of the stability matrix. We then add the contributions of one NNLO operator by random entries in that typical range to the stability matrix, see the Supplemental Material for details. Averaging over $10^5$ sets of random numbers, we obtain an estimate for systematic errors on $\Theta_y$. In \cref{fig:G-Lambda-plot}, we show as a hatched region where between $16\%$ and $84\%$ of random NNLO entries give rise to $\Theta_y>0$. In both cases, the shaded region is confined around the NLO-line and does not overlap with the fixed point.

As a second test, we simulate $\Theta_i$ as a function of the Newton coupling for $\lambda=\lambda^*$, see \cref{fig:ThetasNLOvsNNLO}. Here, we show the upper and lower standard deviations as error bars. We see that while there are still significant error bars on the value of $\Theta_y$, at the fixed point $\Theta_y(g^*)>0$ holds well beyond 1-$\sigma$, see \eqref{eq:theta-Yuk-res}. We further see that the higher-order operators most likely remain irrelevant at NNLO.

Both of these studies indicate that NNLO operators, which we have neglected in this study, likely will not change the qualitative result, namely that $\Theta_y>0$ beyond LO. Furthermore, the simulation of the effect of higher-order operators is a novel method to estimate systematic errors on FRG computations.

\bigskip
%%%%%%%%%%%%%%%%%%%%%%%%%%%%
\textit{Discussion and conclusion---} 
%%%%%%%%%%%%%%%%%%%%%%%%%%%%%%%%%
In this letter, we computed the NLO gravity contribution to the critical exponent of the Yukawa coupling. Within the FRG setup, this NLO contribution arises from a systematic resummation of higher-order operators. We identified two distinct mechanisms by which higher-order operators can contribute. First, induced operators -- which lack a Gau\ss ian fixed point -- become asymptotically safe in the UV and directly modify the leading-order element of the stability matrix. Second, non-induced operators -- which are asymptotically free -- can also affect the critical exponent through off-diagonal elements of the stability matrix. The latter is a novel mechanism that has not been considered so far.

We determined the full set of NLO operators by analysing the diagrams contributing to the Yukawa beta functions, see \cref{fig:Yukawa-flow}. The full set of induced and non-induced operators is listed in \cref{tab:induced,tab:non-induced}, respectively. In the present computation, the non-induced operators provide the dominant contribution, in particular, for Standard Model matter, where the fermion loop mediated by the fermion-kinetic Yukawa coupling is the strongest contribution, see \cref{tab:loop-contribution}.

Our main result is that the Yukawa coupling is relevant in this complete NLO computation. This holds both for minimal matter content ($N_\text{s} = N_\text{f} =1$) and for SM matter, see \cref{fig:ThetasNLOvsNNLO,fig:G-Lambda-plot}. Adding fermions and gauge fields enhances the two-loop coefficient, and enlarges the region in the $g$-$\lambda$ plane where the Yukawa coupling is relevant, while adding scalar fields has a sub-leading effect.

To assess systematic uncertainties, we propose a novel error-estimation method that simulates the influence of NNLO operators. This estimate confirms that the Yukawa coupling lies robustly within the relevant regime. This can be combined with stability studies with respect to gauge, regulator, and projection choices, which is the subject of forthcoming work \cite{longpaper}.

In summary, our work provides both conceptual and technical advances in computing quantum-gravity effects on matter interactions at complete NLO precision. Here, we have analysed the Yukawa system with unprecedented accuracy. While previous computations remained inconclusive, our extended setup indicates that non-vanishing Yukawa couplings are compatible with asymptotically safe quantum gravity, which is a necessary requirement for the Standard Model to emerge from an asymptotically safe ultraviolet completion. Our work sets the basis to study the effect of induced and non-induced NLO operators on the remaining SM couplings in the future. 

\smallskip
%%%%%%%%%%%%%%%%%%%%%%%%%%%%
\textit{Acknowledgements---}
%%%%%%%%%%%%%%%%%%%%%%%%%%%%%%%%%
We thank Astrid Eichhorn, Holger Gies, Zois Gyftopolos, Aaron Held, Benjamin Knorr, Daniel Litim, and Jan Pawlowski for discussions.
GPB is supported by CNPq under the grant PQ-C (308651/2025-1).
MR is supported by the Science and Technology Facilities Council under the Consolidated Grant ST/X000796/1, and the Ernest Rutherford Fellowship ST/Z510282/1. The research of MS was supported by Perimeter Institute during the early stages of this work. The research of MS was also supported by a Radboud Excellence fellowship from Radboud University in Nijmegen, Netherlands, and by an NWO Veni grant under grant ID  [\url{https://doi.org/10.61686/SUPEH07195}]. 
Research at Perimeter Institute is supported in part by the Government of Canada through the Department of Innovation, Science and Economic Development Canada and by the Province of Ontario through the Ministry of Colleges and Universities.

%%%%%% References %%%%%%%%%%%%%%%%%%%%
\bibliographystyle{mystyle}
\bibliography{Gravity}

\providecommand{\href}[2]{#2}\begingroup\raggedright\begin{thebibliography}{10}

\bibitem{Weinberg:1980gg}
S.~Weinberg, \emph{{Ultraviolet divergences in quantum theories of
  gravitation}}, pp.~790--831.
\newblock 1980.

\bibitem{Reuter:1996cp}
M.~Reuter, \emph{{Nonperturbative evolution equation for quantum gravity}},
  \href{https://doi.org/10.1103/PhysRevD.57.971}{\emph{Phys. Rev. D} {\bfseries
  57} (1998) 971} [\href{https://arxiv.org/abs/hep-th/9605030}{{\ttfamily
  hep-th/9605030}}].

\bibitem{Bonanno:2020bil}
A.~Bonanno, A.~Eichhorn, H.~Gies, J.~M. Pawlowski, R.~Percacci, M.~Reuter
  et~al., \emph{{Critical reflections on asymptotically safe gravity}},
  \href{https://doi.org/10.3389/fphy.2020.00269}{\emph{Front. in Phys.}
  {\bfseries 8} (2020) 269} [\href{https://arxiv.org/abs/2004.06810}{{\ttfamily
  2004.06810}}].

\bibitem{Dupuis:2020fhh}
N.~Dupuis, L.~Canet, A.~Eichhorn, W.~Metzner, J.~M. Pawlowski, M.~Tissier
  et~al., \emph{{The nonperturbative functional renormalization group and its
  applications}},
  \href{https://doi.org/10.1016/j.physrep.2021.01.001}{\emph{Phys. Rept.}
  {\bfseries 910} (2021) 1} [\href{https://arxiv.org/abs/2006.04853}{{\ttfamily
  2006.04853}}].

\bibitem{Knorr:2022dsx}
B.~Knorr, C.~Ripken and F.~Saueressig, \emph{{Form Factors in Asymptotically
  Safe Quantum Gravity}},  in \emph{Handbook of Quantum Gravity}, (Singapore),
  Springer Nature Singapore, (2023),
  \href{https://arxiv.org/abs/2210.16072}{{\ttfamily 2210.16072}},
  \href{https://doi.org/10.1007/978-981-19-3079-9_21-1}{DOI}.

\bibitem{Eichhorn:2022gku}
A.~Eichhorn and M.~Schiffer, \emph{{Asymptotic safety of gravity with matter}},
   in \emph{Handbook of Quantum Gravity}, (Singapore), Springer Nature
  Singapore, (12, 2024), \href{https://arxiv.org/abs/2212.07456}{{\ttfamily
  2212.07456}}, \href{https://doi.org/10.1007/978-981-99-7681-2_22}{DOI}.

\bibitem{Morris:2022btf}
T.~R. Morris and D.~Stulga, \emph{{The Functional f(R) Approximation}},  in
  \emph{Handbook of Quantum Gravity}, (Singapore), Springer Nature Singapore,
  (2023), \href{https://arxiv.org/abs/2210.11356}{{\ttfamily 2210.11356}},
  \href{https://doi.org/10.1007/978-981-19-3079-9_19-1}{DOI}.

\bibitem{Wetterich:2022ncl}
C.~Wetterich, \emph{{Quantum Gravity and Scale Symmetry in Cosmology}},  in
  \emph{Handbook of Quantum Gravity}, (Singapore), Springer Nature Singapore,
  (2023), \href{https://arxiv.org/abs/2211.03596}{{\ttfamily 2211.03596}},
  \href{https://doi.org/10.1007/978-981-19-3079-9_26-1}{DOI}.

\bibitem{Martini:2022sll}
R.~Martini, G.~P. Vacca and O.~Zanusso, \emph{{Perturbative Approaches to
  Nonperturbative Quantum Gravity}},  in \emph{Handbook of Quantum Gravity},
  (Singapore), pp.~1--46, Springer Nature Singapore, (2023),
  \href{https://arxiv.org/abs/2210.13910}{{\ttfamily 2210.13910}},
  \href{https://doi.org/10.1007/978-981-19-3079-9_25-1}{DOI}.

\bibitem{Saueressig:2023irs}
F.~Saueressig, \emph{{The Functional Renormalization Group in Quantum
  Gravity}},  in \emph{Handbook of Quantum Gravity}, (Singapore), pp.~1--33,
  Springer Nature Singapore, (2023),
  \href{https://arxiv.org/abs/2302.14152}{{\ttfamily 2302.14152}},
  \href{https://doi.org/10.1007/978-981-19-3079-9_16-1}{DOI}.

\bibitem{Pawlowski:2023gym}
J.~M. Pawlowski and M.~Reichert, \emph{{Quantum Gravity from dynamical metric
  fluctuations}},  in \emph{Handbook of Quantum Gravity}, (Singapore), Springer
  Nature Singapore, (9, 2023),
  \href{https://arxiv.org/abs/2309.10785}{{\ttfamily 2309.10785}},
  \href{https://doi.org/10.1007/978-981-19-3079-9_17-1}{DOI}.

\bibitem{Platania:2023srt}
A.~Platania, \emph{Black holes in asymptotically safe gravity},  in
  \emph{Handbook of Quantum Gravity}, (Singapore), Springer Nature Singapore,
  (2023), \href{https://arxiv.org/abs/2302.04272}{{\ttfamily 2302.04272}},
  \href{https://doi.org/10.1007/978-981-19-3079-9_24-1}{DOI}.

\bibitem{Bonanno:2024xne}
A.~Bonanno, \emph{{Asymptotic Safety and Cosmology}},  in \emph{Handbook of
  Quantum Gravity}, (Singapore), Springer Nature Singapore, (2023),
  \href{https://doi.org/10.1007/978-981-19-3079-9_23-1}{DOI}.

\bibitem{Reichert:2020mja}
M.~Reichert, \emph{{Lecture notes: Functional Renormalisation Group and
  Asymptotically Safe Quantum Gravity}},
  \href{https://doi.org/10.22323/1.384.0005}{\emph{PoS} {\bfseries 384} (2020)
  005}.

\bibitem{Basile:2024oms}
I.~Basile, L.~Buoninfante, F.~Di~Filippo, B.~Knorr, A.~Platania and
  A.~Tokareva, \emph{{Lectures in quantum gravity}},
  \href{https://doi.org/10.21468/SciPostPhysLectNotes.98}{\emph{SciPost Phys.
  Lect. Notes} {\bfseries 98} (2025) 1}
  [\href{https://arxiv.org/abs/2412.08690}{{\ttfamily 2412.08690}}].

\bibitem{Dona:2013qba}
P.~Don\`a, A.~Eichhorn and R.~Percacci, \emph{{Matter matters in asymptotically
  safe quantum gravity}},
  \href{https://doi.org/10.1103/PhysRevD.89.084035}{\emph{Phys. Rev. D}
  {\bfseries 89} (2014) 084035}
  [\href{https://arxiv.org/abs/1311.2898}{{\ttfamily 1311.2898}}].

\bibitem{Meibohm:2015twa}
J.~Meibohm, J.~M. Pawlowski and M.~Reichert, \emph{{Asymptotic safety of
  gravity-matter systems}},
  \href{https://doi.org/10.1103/PhysRevD.93.084035}{\emph{Phys. Rev. D}
  {\bfseries 93} (2016) 084035}
  [\href{https://arxiv.org/abs/1510.07018}{{\ttfamily 1510.07018}}].

\bibitem{Biemans:2017zca}
J.~Biemans, A.~Platania and F.~Saueressig, \emph{{Renormalization group fixed
  points of foliated gravity-matter systems}},
  \href{https://doi.org/10.1007/JHEP05(2017)093}{\emph{JHEP} {\bfseries 05}
  (2017) 093} [\href{https://arxiv.org/abs/1702.06539}{{\ttfamily
  1702.06539}}].

\bibitem{Christiansen:2017cxa}
N.~Christiansen, D.~F. Litim, J.~M. Pawlowski and M.~Reichert,
  \emph{{Asymptotic safety of gravity with matter}},
  \href{https://doi.org/10.1103/PhysRevD.97.106012}{\emph{Phys. Rev. D}
  {\bfseries 97} (2018) 106012}
  [\href{https://arxiv.org/abs/1710.04669}{{\ttfamily 1710.04669}}].

\bibitem{Alkofer:2018fxj}
N.~Alkofer and F.~Saueressig, \emph{{Asymptotically safe $f(R)$-gravity coupled
  to matter I: the polynomial case}},
  \href{https://doi.org/10.1016/j.aop.2018.07.017}{\emph{Annals Phys.}
  {\bfseries 396} (2018) 173}
  [\href{https://arxiv.org/abs/1802.00498}{{\ttfamily 1802.00498}}].

\bibitem{Eichhorn:2018akn}
A.~Eichhorn, P.~Labus, J.~M. Pawlowski and M.~Reichert, \emph{{Effective
  universality in quantum gravity}},
  \href{https://doi.org/10.21468/SciPostPhys.5.4.031}{\emph{SciPost Phys.}
  {\bfseries 5} (2018) 031} [\href{https://arxiv.org/abs/1804.00012}{{\ttfamily
  1804.00012}}].

\bibitem{Eichhorn:2018ydy}
A.~Eichhorn, S.~Lippoldt, J.~M. Pawlowski, M.~Reichert and M.~Schiffer,
  \emph{{How perturbative is quantum gravity?}},
  \href{https://doi.org/10.1016/j.physletb.2019.01.071}{\emph{Phys. Lett. B}
  {\bfseries 792} (2019) 310}
  [\href{https://arxiv.org/abs/1810.02828}{{\ttfamily 1810.02828}}].

\bibitem{Eichhorn:2018nda}
A.~Eichhorn, S.~Lippoldt and M.~Schiffer, \emph{{Zooming in on fermions and
  quantum gravity}},
  \href{https://doi.org/10.1103/PhysRevD.99.086002}{\emph{Phys. Rev. D}
  {\bfseries 99} (2019) 086002}
  [\href{https://arxiv.org/abs/1812.08782}{{\ttfamily 1812.08782}}].

\bibitem{Burger:2019upn}
B.~B\"urger, J.~M. Pawlowski, M.~Reichert and B.-J. Schaefer, \emph{{Curvature
  dependence of quantum gravity with scalars}},
  \href{https://arxiv.org/abs/1912.01624}{{\ttfamily 1912.01624}}.

\bibitem{Wetterich:2019zdo}
C.~Wetterich and M.~Yamada, \emph{{Variable Planck mass from the gauge
  invariant flow equation}},
  \href{https://doi.org/10.1103/PhysRevD.100.066017}{\emph{Phys.\ Rev.\ D}
  {\bfseries 100} (2019) 066017}
  [\href{https://arxiv.org/abs/1906.01721}{{\ttfamily 1906.01721}}].

\bibitem{deBrito:2020xhy}
G.~P. de~Brito, A.~D. Pereira and A.~F. Vieira, \emph{{Exploring new corners of
  asymptotically safe unimodular quantum gravity}},
  \href{https://doi.org/10.1103/PhysRevD.103.104023}{\emph{Phys. Rev. D}
  {\bfseries 103} (2021) 104023}
  [\href{https://arxiv.org/abs/2012.08904}{{\ttfamily 2012.08904}}].

\bibitem{Pastor-Gutierrez:2022nki}
A.~Pastor-Guti\'errez, J.~M. Pawlowski and M.~Reichert, \emph{{The
  Asymptotically Safe Standard Model: From quantum gravity to dynamical chiral
  symmetry breaking}},
  \href{https://doi.org/10.21468/SciPostPhys.15.3.105}{\emph{SciPost Phys.}
  {\bfseries 15} (2023) 105}
  [\href{https://arxiv.org/abs/2207.09817}{{\ttfamily 2207.09817}}].

\bibitem{Korver:2024sam}
G.~Korver, F.~Saueressig and J.~Wang, \emph{{Global flows of foliated
  gravity-matter systems}},
  \href{https://doi.org/10.1016/j.physletb.2024.138789}{\emph{Phys. Lett. B}
  {\bfseries 855} (2024) 138789}
  [\href{https://arxiv.org/abs/2402.01260}{{\ttfamily 2402.01260}}].

\bibitem{Eichhorn:2017ylw}
A.~Eichhorn and A.~Held, \emph{{Top mass from asymptotic safety}},
  \href{https://doi.org/10.1016/j.physletb.2017.12.040}{\emph{Phys. Lett. B}
  {\bfseries 777} (2018) 217}
  [\href{https://arxiv.org/abs/1707.01107}{{\ttfamily 1707.01107}}].

\bibitem{Alkofer:2020vtb}
R.~Alkofer, A.~Eichhorn, A.~Held, C.~M. Nieto, R.~Percacci and M.~Schr\"ofl,
  \emph{{Quark masses and mixings in minimally parameterized UV completions of
  the Standard Model}},
  \href{https://doi.org/10.1016/j.aop.2020.168282}{\emph{Annals Phys.}
  {\bfseries 421} (2020) 168282}
  [\href{https://arxiv.org/abs/2003.08401}{{\ttfamily 2003.08401}}].

\bibitem{Eichhorn:2025sux}
A.~Eichhorn, Z.~Gyftopoulos and A.~Held, \emph{{Quark and lepton mixing in the
  asymptotically safe Standard Model}},
  \href{https://arxiv.org/abs/2507.18304}{{\ttfamily 2507.18304}}.

\bibitem{Eichhorn:2018whv}
A.~Eichhorn and A.~Held, \emph{{Mass difference for charged quarks from
  asymptotically safe quantum gravity}},
  \href{https://doi.org/10.1103/PhysRevLett.121.151302}{\emph{Phys. Rev. Lett.}
  {\bfseries 121} (2018) 151302}
  [\href{https://arxiv.org/abs/1803.04027}{{\ttfamily 1803.04027}}].

\bibitem{Shaposhnikov:2009pv}
M.~Shaposhnikov and C.~Wetterich, \emph{{Asymptotic safety of gravity and the
  Higgs boson mass}},
  \href{https://doi.org/10.1016/j.physletb.2009.12.022}{\emph{Phys. Lett. B}
  {\bfseries 683} (2010) 196}
  [\href{https://arxiv.org/abs/0912.0208}{{\ttfamily 0912.0208}}].

\bibitem{Daum:2009dn}
J.-E. Daum, U.~Harst and M.~Reuter, \emph{{Running Gauge Coupling in
  Asymptotically Safe Quantum Gravity}},
  \href{https://doi.org/10.1007/JHEP01(2010)084}{\emph{JHEP} {\bfseries 1001}
  (2010) 084} [\href{https://arxiv.org/abs/0910.4938}{{\ttfamily 0910.4938}}].

\bibitem{Harst:2011zx}
U.~Harst and M.~Reuter, \emph{{QED coupled to QEG}},
  \href{https://doi.org/10.1007/JHEP05(2011)119}{\emph{JHEP} {\bfseries 05}
  (2011) 119} [\href{https://arxiv.org/abs/1101.6007}{{\ttfamily 1101.6007}}].

\bibitem{Folkerts:2011jz}
S.~Folkerts, D.~F. Litim and J.~M. Pawlowski, \emph{{Asymptotic freedom of
  Yang-Mills theory with gravity}},
  \href{https://doi.org/10.1016/j.physletb.2012.02.002}{\emph{Phys. Lett. B}
  {\bfseries 709} (2012) 234}
  [\href{https://arxiv.org/abs/1101.5552}{{\ttfamily 1101.5552}}].

\bibitem{Christiansen:2017gtg}
N.~Christiansen and A.~Eichhorn, \emph{{An asymptotically safe solution to the
  U(1) triviality problem}},
  \href{https://doi.org/10.1016/j.physletb.2017.04.047}{\emph{Phys. Lett. B}
  {\bfseries 770} (2017) 154}
  [\href{https://arxiv.org/abs/1702.07724}{{\ttfamily 1702.07724}}].

\bibitem{Eichhorn:2017lry}
A.~Eichhorn and F.~Versteegen, \emph{{Upper bound on the Abelian gauge coupling
  from asymptotic safety}},
  \href{https://doi.org/10.1007/JHEP01(2018)030}{\emph{JHEP} {\bfseries 01}
  (2018) 030} [\href{https://arxiv.org/abs/1709.07252}{{\ttfamily
  1709.07252}}].

\bibitem{Eichhorn:2019yzm}
A.~Eichhorn and M.~Schiffer, \emph{{$d=4$ as the critical dimensionality of
  asymptotically safe interactions}},
  \href{https://doi.org/10.1016/j.physletb.2019.05.005}{\emph{Phys. Lett. B}
  {\bfseries 793} (2019) 383}
  [\href{https://arxiv.org/abs/1902.06479}{{\ttfamily 1902.06479}}].

\bibitem{deBrito:2019umw}
G.~P. De~Brito, A.~Eichhorn and A.~D. Pereira, \emph{{A link that matters:
  Towards phenomenological tests of unimodular asymptotic safety}},
  \href{https://doi.org/10.1007/JHEP09(2019)100}{\emph{JHEP} {\bfseries 09}
  (2019) 100} [\href{https://arxiv.org/abs/1907.11173}{{\ttfamily
  1907.11173}}].

\bibitem{deBrito:2022vbr}
G.~P. de~Brito and A.~Eichhorn, \emph{{Nonvanishing gravitational contribution
  to matter beta functions for vanishing dimensionful regulators}},
  \href{https://doi.org/10.1140/epjc/s10052-023-11172-z}{\emph{Eur. Phys. J. C}
  {\bfseries 83} (2023) 161}
  [\href{https://arxiv.org/abs/2201.11402}{{\ttfamily 2201.11402}}].

\bibitem{Riabokon:2025ozw}
M.~Riabokon, M.~Schiffer and F.~Wagner, \emph{{Regulator and gauge dependence
  of the Abelian gauge coupling in asymptotically safe quantum gravity}},
  \href{https://arxiv.org/abs/2508.03563}{{\ttfamily 2508.03563}}.

\bibitem{Narain:2009fy}
G.~Narain and R.~Percacci, \emph{{Renormalization Group Flow in Scalar-Tensor
  Theories. I}},
  \href{https://doi.org/10.1088/0264-9381/27/7/075001}{\emph{Class. Quant.
  Grav.} {\bfseries 27} (2010) 075001}
  [\href{https://arxiv.org/abs/0911.0386}{{\ttfamily 0911.0386}}].

\bibitem{Percacci:2015wwa}
R.~Percacci and G.~P. Vacca, \emph{{Search of scaling solutions in
  scalar-tensor gravity}},
  \href{https://doi.org/10.1140/epjc/s10052-015-3410-0}{\emph{Eur. Phys. J.}
  {\bfseries C75} (2015) 188}
  [\href{https://arxiv.org/abs/1501.00888}{{\ttfamily 1501.00888}}].

\bibitem{Labus:2015ska}
P.~Labus, R.~Percacci and G.~P. Vacca, \emph{{Asymptotic safety in O(N) scalar
  models coupled to gravity}},
  \href{https://doi.org/10.1016/j.physletb.2015.12.022}{\emph{Phys. Lett.}
  {\bfseries B753} (2016) 274}
  [\href{https://arxiv.org/abs/1505.05393}{{\ttfamily 1505.05393}}].

\bibitem{Oda:2015sma}
K.-y. Oda and M.~Yamada, \emph{{Non-minimal coupling in Higgs--Yukawa model
  with asymptotically safe gravity}},
  \href{https://doi.org/10.1088/0264-9381/33/12/125011}{\emph{Class. Quant.
  Grav.} {\bfseries 33} (2016) 125011}
  [\href{https://arxiv.org/abs/1510.03734}{{\ttfamily 1510.03734}}].

\bibitem{Hamada:2017rvn}
Y.~Hamada and M.~Yamada, \emph{{Asymptotic safety of higher derivative quantum
  gravity non-minimally coupled with a matter system}},
  \href{https://doi.org/10.1007/JHEP08(2017)070}{\emph{JHEP} {\bfseries 08}
  (2017) 070} [\href{https://arxiv.org/abs/1703.09033}{{\ttfamily
  1703.09033}}].

\bibitem{Eichhorn:2017als}
A.~Eichhorn, Y.~Hamada, J.~Lumma and M.~Yamada, \emph{{Quantum gravity
  fluctuations flatten the Planck-scale Higgs potential}},
  \href{https://doi.org/10.1103/PhysRevD.97.086004}{\emph{Phys. Rev.}
  {\bfseries D97} (2018) 086004}
  [\href{https://arxiv.org/abs/1712.00319}{{\ttfamily 1712.00319}}].

\bibitem{Pawlowski:2018ixd}
J.~M. Pawlowski, M.~Reichert, C.~Wetterich and M.~Yamada, \emph{{Higgs scalar
  potential in asymptotically safe quantum gravity}},
  \href{https://doi.org/10.1103/PhysRevD.99.086010}{\emph{Phys.\ Rev.\ D}
  {\bfseries 99} (2019) 086010}
  [\href{https://arxiv.org/abs/1811.11706}{{\ttfamily 1811.11706}}].

\bibitem{Wetterich:2019rsn}
C.~Wetterich, \emph{{Effective scalar potential in asymptotically safe quantum
  gravity}}, \href{https://doi.org/10.3390/universe7020045}{\emph{Universe}
  {\bfseries 7} (2021) 45} [\href{https://arxiv.org/abs/1911.06100}{{\ttfamily
  1911.06100}}].

\bibitem{Eichhorn:2020sbo}
A.~Eichhorn and M.~Pauly, \emph{{Constraining power of asymptotic safety for
  scalar fields}},
  \href{https://doi.org/10.1103/PhysRevD.103.026006}{\emph{Phys. Rev. D}
  {\bfseries 103} (2021) 026006}
  [\href{https://arxiv.org/abs/2009.13543}{{\ttfamily 2009.13543}}].

\bibitem{Zanusso:2009bs}
O.~Zanusso, L.~Zambelli, G.~P. Vacca and R.~Percacci, \emph{{Gravitational
  corrections to Yukawa systems}},
  \href{https://doi.org/10.1016/j.physletb.2010.04.043}{\emph{Phys. Lett.}
  {\bfseries B689} (2010) 90}
  [\href{https://arxiv.org/abs/0904.0938}{{\ttfamily 0904.0938}}].

\bibitem{Eichhorn:2016esv}
A.~Eichhorn, A.~Held and J.~M. Pawlowski, \emph{{Quantum-gravity effects on a
  Higgs-Yukawa model}},
  \href{https://doi.org/10.1103/PhysRevD.94.104027}{\emph{Phys. Rev. D}
  {\bfseries 94} (2016) 104027}
  [\href{https://arxiv.org/abs/1604.02041}{{\ttfamily 1604.02041}}].

\bibitem{Eichhorn:2017eht}
A.~Eichhorn and A.~Held, \emph{{Viability of quantum-gravity induced
  ultraviolet completions for matter}},
  \href{https://doi.org/10.1103/PhysRevD.96.086025}{\emph{Phys. Rev. D}
  {\bfseries 96} (2017) 086025}
  [\href{https://arxiv.org/abs/1705.02342}{{\ttfamily 1705.02342}}].

\bibitem{Rodigast:2009zj}
A.~Rodigast and T.~Schuster, \emph{{Gravitational Corrections to Yukawa and
  phi**4 Interactions}},
  \href{https://doi.org/10.1103/PhysRevLett.104.081301}{\emph{Phys. Rev. Lett.}
  {\bfseries 104} (2010) 081301}
  [\href{https://arxiv.org/abs/0908.2422}{{\ttfamily 0908.2422}}].

\bibitem{Wetterich:1992yh}
C.~Wetterich, \emph{{Exact evolution equation for the effective potential}},
  \href{https://doi.org/10.1016/0370-2693(93)90726-X}{\emph{Phys. Lett. B}
  {\bfseries 301} (1993) 90}
  [\href{https://arxiv.org/abs/1710.05815}{{\ttfamily 1710.05815}}].

\bibitem{Morris:1993qb}
T.~R. Morris, \emph{{The Exact renormalization group and approximate
  solutions}}, \href{https://doi.org/10.1142/S0217751X94000972}{\emph{Int. J.
  Mod. Phys. A} {\bfseries 9} (1994) 2411}
  [\href{https://arxiv.org/abs/hep-ph/9308265}{{\ttfamily hep-ph/9308265}}].

\bibitem{Ellwanger:1993mw}
U.~Ellwanger, \emph{{FLow equations for N point functions and bound states}},
  \href{https://doi.org/10.1007/BF01555911}{\emph{Z. Phys. C} {\bfseries 62}
  (1994) 503} [\href{https://arxiv.org/abs/hep-ph/9308260}{{\ttfamily
  hep-ph/9308260}}].

\bibitem{Eichhorn:2017sok}
A.~Eichhorn, S.~Lippoldt and V.~Skrinjar, \emph{{Nonminimal hints for
  asymptotic safety}},
  \href{https://doi.org/10.1103/PhysRevD.97.026002}{\emph{Phys. Rev. D}
  {\bfseries 97} (2018) 026002}
  [\href{https://arxiv.org/abs/1710.03005}{{\ttfamily 1710.03005}}].

\bibitem{Laporte:2021kyp}
C.~Laporte, A.~D. Pereira, F.~Saueressig and J.~Wang, \emph{{Scalar-tensor
  theories within Asymptotic Safety}},
  \href{https://doi.org/10.1007/JHEP12(2021)001}{\emph{JHEP} {\bfseries 12}
  (2021) 001} [\href{https://arxiv.org/abs/2110.09566}{{\ttfamily
  2110.09566}}].

\bibitem{Eichhorn:2011pc}
A.~Eichhorn and H.~Gies, \emph{{Light fermions in quantum gravity}},
  \href{https://doi.org/10.1088/1367-2630/13/12/125012}{\emph{New J. Phys.}
  {\bfseries 13} (2011) 125012}
  [\href{https://arxiv.org/abs/1104.5366}{{\ttfamily 1104.5366}}].

\bibitem{Meibohm:2016mkp}
J.~Meibohm and J.~M. Pawlowski, \emph{{Chiral fermions in asymptotically safe
  quantum gravity}},
  \href{https://doi.org/10.1140/epjc/s10052-016-4132-7}{\emph{Eur. Phys. J. C}
  {\bfseries 76} (2016) 285}
  [\href{https://arxiv.org/abs/1601.04597}{{\ttfamily 1601.04597}}].

\bibitem{Eichhorn:2012va}
A.~Eichhorn, \emph{{Quantum-gravity-induced matter self-interactions in the
  asymptotic-safety scenario}},
  \href{https://doi.org/10.1103/PhysRevD.86.105021}{\emph{Phys. Rev.}
  {\bfseries D86} (2012) 105021}
  [\href{https://arxiv.org/abs/1204.0965}{{\ttfamily 1204.0965}}].

\bibitem{Knorr:2022ilz}
B.~Knorr, \emph{{Safe essential scalar-tensor theories}},
  \href{https://arxiv.org/abs/2204.08564}{{\ttfamily 2204.08564}}.

\bibitem{deBrito:2023myf}
G.~P. de~Brito, B.~Knorr and M.~Schiffer, \emph{{On the weak-gravity bound for
  a shift-symmetric scalar field}},
  \href{https://doi.org/10.1103/PhysRevD.108.026004}{\emph{Phys. Rev. D}
  {\bfseries 108} (2023) 026004}
  [\href{https://arxiv.org/abs/2302.10989}{{\ttfamily 2302.10989}}].

\bibitem{deBrito:2019epw}
G.~P. De~Brito, Y.~Hamada, A.~D. Pereira and M.~Yamada, \emph{{On the impact of
  Majorana masses in gravity-matter systems}},
  \href{https://doi.org/10.1007/JHEP08(2019)142}{\emph{JHEP} {\bfseries 08}
  (2019) 142} [\href{https://arxiv.org/abs/1905.11114}{{\ttfamily
  1905.11114}}].

\bibitem{deBrito:2020dta}
G.~P. de~Brito, A.~Eichhorn and M.~Schiffer, \emph{{Light charged fermions in
  quantum gravity}},
  \href{https://doi.org/10.1016/j.physletb.2021.136128}{\emph{Phys. Lett. B}
  {\bfseries 815} (2021) 136128}
  [\href{https://arxiv.org/abs/2010.00605}{{\ttfamily 2010.00605}}].

\bibitem{Eichhorn:2021qet}
A.~Eichhorn, J.~H. Kwapisz and M.~Schiffer, \emph{{Weak-gravity bound in
  asymptotically safe gravity-gauge systems}},
  \href{https://doi.org/10.1103/PhysRevD.105.106022}{\emph{Phys. Rev. D}
  {\bfseries 105} (2022) 106022}
  [\href{https://arxiv.org/abs/2112.09772}{{\ttfamily 2112.09772}}].

\bibitem{deBrito:2021pyi}
G.~P. de~Brito, A.~Eichhorn and R.~R. L.~d. Santos, \emph{{The weak-gravity
  bound and the need for spin in asymptotically safe matter-gravity models}},
  \href{https://doi.org/10.1007/JHEP11(2021)110}{\emph{JHEP} {\bfseries 11}
  (2021) 110} [\href{https://arxiv.org/abs/2107.03839}{{\ttfamily
  2107.03839}}].

\bibitem{Falls:2013bv}
K.~Falls, D.~Litim, K.~Nikolakopoulos and C.~Rahmede, \emph{{A bootstrap
  towards asymptotic safety}},
  \href{https://arxiv.org/abs/1301.4191}{{\ttfamily 1301.4191}}.

\bibitem{Falls:2014tra}
K.~Falls, D.~F. Litim, K.~Nikolakopoulos and C.~Rahmede, \emph{{Further
  evidence for asymptotic safety of quantum gravity}},
  \href{https://doi.org/10.1103/PhysRevD.93.104022}{\emph{Phys. Rev.}
  {\bfseries D93} (2016) 104022}
  [\href{https://arxiv.org/abs/1410.4815}{{\ttfamily 1410.4815}}].

\bibitem{Denz:2016qks}
T.~Denz, J.~M. Pawlowski and M.~Reichert, \emph{{Towards apparent convergence
  in asymptotically safe quantum gravity}},
  \href{https://doi.org/10.1140/epjc/s10052-018-5806-0}{\emph{Eur. Phys. J. C}
  {\bfseries 78} (2018) 336}
  [\href{https://arxiv.org/abs/1612.07315}{{\ttfamily 1612.07315}}].

\bibitem{Falls:2017lst}
K.~Falls, C.~R. King, D.~F. Litim, K.~Nikolakopoulos and C.~Rahmede,
  \emph{{Asymptotic safety of quantum gravity beyond Ricci scalars}},
  \href{https://doi.org/10.1103/PhysRevD.97.086006}{\emph{Phys. Rev.}
  {\bfseries D97} (2018) 086006}
  [\href{https://arxiv.org/abs/1801.00162}{{\ttfamily 1801.00162}}].

\bibitem{Falls:2018ylp}
K.~G. Falls, D.~F. Litim and J.~Schröder, \emph{{Aspects of asymptotic safety
  for quantum gravity}},
  \href{https://doi.org/10.1103/PhysRevD.99.126015}{\emph{Phys.\ Rev.\ D}
  {\bfseries 99} (2019) 126015}
  [\href{https://arxiv.org/abs/1810.08550}{{\ttfamily 1810.08550}}].

\bibitem{Kluth:2020bdv}
Y.~Kluth and D.~F. Litim, \emph{{Fixed points of quantum gravity and the
  dimensionality of the UV critical surface}},
  \href{https://doi.org/10.1103/PhysRevD.108.026005}{\emph{Phys. Rev. D}
  {\bfseries 108} (2023) 026005}
  [\href{https://arxiv.org/abs/2008.09181}{{\ttfamily 2008.09181}}].

\bibitem{longpaper}
G.~P. de~Brito, M.~Reichert and M.~Schiffer, \emph{in preparation}, .

\bibitem{Christiansen:2012rx}
N.~Christiansen, D.~F. Litim, J.~M. Pawlowski and A.~Rodigast, \emph{{Fixed
  points and infrared completion of quantum gravity}},
  \href{https://doi.org/10.1016/j.physletb.2013.11.025}{\emph{Phys. Lett. B}
  {\bfseries 728} (2014) 114}
  [\href{https://arxiv.org/abs/1209.4038}{{\ttfamily 1209.4038}}].

\bibitem{Christiansen:2014raa}
N.~Christiansen, B.~Knorr, J.~M. Pawlowski and A.~Rodigast, \emph{{Global Flows
  in Quantum Gravity}},
  \href{https://doi.org/10.1103/PhysRevD.93.044036}{\emph{Phys. Rev. D}
  {\bfseries 93} (2016) 044036}
  [\href{https://arxiv.org/abs/1403.1232}{{\ttfamily 1403.1232}}].

\bibitem{Christiansen:2015rva}
N.~Christiansen, B.~Knorr, J.~Meibohm, J.~M. Pawlowski and M.~Reichert,
  \emph{{Local Quantum Gravity}},
  \href{https://doi.org/10.1103/PhysRevD.92.121501}{\emph{Phys. Rev. D}
  {\bfseries 92} (2015) 121501}
  [\href{https://arxiv.org/abs/1506.07016}{{\ttfamily 1506.07016}}].

\bibitem{Pawlowski:2020qer}
J.~M. Pawlowski and M.~Reichert, \emph{{Quantum Gravity: A Fluctuating Point of
  View}}, \href{https://doi.org/10.3389/fphy.2020.551848}{\emph{Front. in
  Phys.} {\bfseries 8} (2021) 551848}
  [\href{https://arxiv.org/abs/2007.10353}{{\ttfamily 2007.10353}}].

\bibitem{deBrito:2023kow}
G.~P. de~Brito, A.~Eichhorn and S.~Ray, \emph{{Why the quark mass is not the
  Planck mass}}, \href{https://doi.org/10.1103/4jzf-byxc}{\emph{Phys. Rev. D}
  {\bfseries 112} (2025) 046013}
  [\href{https://arxiv.org/abs/2311.16066}{{\ttfamily 2311.16066}}].

\bibitem{AScodebase}
\href{https://as-codebase.quantum-spacetime.net/}{https://as-codebase.quantum-spacetime.net/}.

\bibitem{Brizuela:2008ra}
D.~Brizuela, J.~M. Martin-Garcia and G.~A. Mena~Marugan, \emph{{xPert: Computer
  algebra for metric perturbation theory}},
  \href{https://doi.org/10.1007/s10714-009-0773-2}{\emph{Gen. Rel. Grav.}
  {\bfseries 41} (2009) 2415}
  [\href{https://arxiv.org/abs/0807.0824}{{\ttfamily 0807.0824}}].

\bibitem{Martin-Garcia:2007bqa}
J.~M. Martin-Garcia, R.~Portugal and L.~R.~U. Manssur, \emph{{The Invar tensor
  package}}, \href{https://doi.org/10.1016/j.cpc.2007.05.015}{\emph{Comput.
  Phys. Commun.} {\bfseries 177} (2007) 640}
  [\href{https://arxiv.org/abs/0704.1756}{{\ttfamily 0704.1756}}].

\bibitem{Martin-Garcia:2008yei}
J.~M. Martin-Garcia, D.~Yllanes and R.~Portugal, \emph{{The Invar tensor
  package: Differential invariants of Riemann}},
  \href{https://doi.org/10.1016/j.cpc.2008.04.018}{\emph{Comput. Phys. Commun.}
  {\bfseries 179} (2008) 586}
  [\href{https://arxiv.org/abs/0802.1274}{{\ttfamily 0802.1274}}].

\bibitem{Martin-Garcia:2008ysv}
J.~M. Mart\'\i{}n-Garc\'\i{}a, \emph{{xPerm: fast index canonicalization for
  tensor computer algebra}},
  \href{https://doi.org/10.1016/j.cpc.2008.05.009}{\emph{Comput. Phys. Commun.}
  {\bfseries 179} (2008) 597}
  [\href{https://arxiv.org/abs/0803.0862}{{\ttfamily 0803.0862}}].

\bibitem{Cyrol:2016zqb}
A.~K. Cyrol, M.~Mitter and N.~Strodthoff, \emph{{FormTracer - A Mathematica
  Tracing Package Using FORM}},
  \href{https://doi.org/10.1016/j.cpc.2017.05.024}{\emph{Comput. Phys. Commun.}
  {\bfseries 219} (2017) 346}
  [\href{https://arxiv.org/abs/1610.09331}{{\ttfamily 1610.09331}}].

\bibitem{Huber:2011qr}
M.~Q. Huber and J.~Braun, \emph{{Algorithmic derivation of functional
  renormalization group equations and Dyson-Schwinger equations}},
  \href{https://doi.org/10.1016/j.cpc.2012.01.014}{\emph{Comput.\ Phys.\
  Commun.} {\bfseries 183} (2012) 1290}
  [\href{https://arxiv.org/abs/1102.5307}{{\ttfamily 1102.5307}}].

\bibitem{Huber:2019dkb}
M.~Q. Huber, A.~K. Cyrol and J.~M. Pawlowski, \emph{{DoFun 3.0: Functional
  equations in Mathematica}},
  \href{https://doi.org/10.1016/j.cpc.2019.107058}{\emph{Comput. Phys. Commun.}
  {\bfseries 248} (2020) 107058}
  [\href{https://arxiv.org/abs/1908.02760}{{\ttfamily 1908.02760}}].

\bibitem{Vermaseren:2000nd}
J.~A.~M. Vermaseren, \emph{{New features of FORM}},
  \href{https://arxiv.org/abs/math-ph/0010025}{{\ttfamily math-ph/0010025}}.

\bibitem{Kuipers:2012rf}
J.~Kuipers, T.~Ueda, J.~A.~M. Vermaseren and J.~Vollinga, \emph{{FORM version
  4.0}}, \href{https://doi.org/10.1016/j.cpc.2012.12.028}{\emph{Comput. Phys.
  Commun.} {\bfseries 184} (2013) 1453}
  [\href{https://arxiv.org/abs/1203.6543}{{\ttfamily 1203.6543}}].

\bibitem{Litim:2000ci}
D.~F. Litim, \emph{{Optimization of the exact renormalization group}},
  \href{https://doi.org/10.1016/S0370-2693(00)00748-6}{\emph{Phys.Lett.}
  {\bfseries B486} (2000) 92}
  [\href{https://arxiv.org/abs/hep-th/0005245}{{\ttfamily hep-th/0005245}}].

\bibitem{Balog:2019rrg}
I.~Balog, H.~Chat{\'e}, B.~Delamotte, M.~Marohnic and N.~Wschebor,
  \emph{{Convergence of Nonperturbative Approximations to the Renormalization
  Group}}, \href{https://doi.org/10.1103/PhysRevLett.123.240604}{\emph{Phys.
  Rev. Lett.} {\bfseries 123} (2019) 240604}
  [\href{https://arxiv.org/abs/1907.01829}{{\ttfamily 1907.01829}}].

\end{thebibliography}\endgroup

%%%%%%%%%%%%%%%%%%%%%%%%%%%%%%%%%%%%%%
%%%%%%%%%%%%%%%%%%%%%%%%%%%%%%%%%%%%%%

\clearpage

%\appendix 
\renewcommand{\thesubsection}{{S.\arabic{subsection}}}
\setcounter{section}{0}

\onecolumngrid

%%%%%%%%%%%%%%%%%%%%%%%%%%%%%%%%%
\section*{Supplemental material}
%%%%%%%%%%%%%%%%%%%%%%%%%%%%%%%%%

In this supplement, we provide technical details omitted in the main text.

\subsection{General setup}

In the current work, we consider gravity coupled to a set of matter fields, including: $N_\text{s}$ scalars $\phi_I$ (with $I = 1,\,2,\,\ldots,\,N_\text{s}$), $N_\text{f}$ Dirac fermions $\psi_i$ (with $i = 1,\,2,\,\ldots,\,N_\text{f}$), and $N_\text{v}$ vectors $A_\mu^a$ (with $a = 1,\,2,\,\ldots,\,N_\text{v}$). We approximate the coarse-grained dynamics of the system with the following ansatz for the scale-dependent effective action $\Gamma_k$,
\begin{align}\label{eq:truncation_full}
\Gamma_k=\Gamma_k^{\mathrm{Grav}}+\Gamma_k^{\mathrm{Matter}}+\Gamma_k^{\mathrm{Induced}}+\Gamma_k^{\mathrm{Yukawa}}\,.
\end{align}
The first term approximates the gravitational dynamics with the Einstein-Hilbert truncation,
\begin{align}
\Gamma_k^{\mathrm{Grav}}=\frac{1}{16\pi\,\GNewton} \int \mathrm d^4 x \sqrt{g} \left(2\CC - R \right)\,
+ \Gamma_k^{\mathrm{gf}}\,,
\end{align}
where $\Gamma_k^{\mathrm{gf}}$ includes standard gauge-fixing terms and the corresponding Faddeev-Popov action. We work in a de-Donder-type harmonic Landau gauge. As usual, we define the dimensionless couplings $\gnewton =k^2 \GNewton$ and $\lambda = k^{-2} \CC$, see \cref{eq:dimless-couplings}. 

We compute correlation functions involving the metric fluctuation $h_{\mu\nu}$, which is defined in terms of the linear metric split 
\begin{align}
g_{\mu\nu} = \delta_{\mu\nu} + \sqrt{16\pi \GNewton Z_h}\, h_{\mu\nu} \,,
\end{align}
where $Z_h$ is the wave function renormalization factor for the fluctuation field $h_{\mu\nu}$ and $\delta_{\mu\nu}$ is the (Euclidean) flat metric.

The second term in \cref{eq:truncation_full} includes matter fields minimally coupled to gravity,
\begin{align}
\Gamma_k^{\mathrm{Matter}} = 
\int \mathrm d^4 x \sqrt{g}\left( \frac{Z_\phi}{2} \partial_\mu \phi_I  \partial^\mu \phi_I +Z_\psi \,i \bar{\psi}_i \slashed{D} \psi_i  +\frac{Z_A}{4} F_{\mu\nu}^a F^{a \mu\nu} \right) ,
\end{align}
where $F_{\mu\nu}^{a} = \partial_\mu A_\nu^a - \partial_\nu A_\mu^a$ and $Z_\phi$, $Z_\psi$ and $Z_A$ are the wave function renormalisation factors for the corresponding matter fields. We also supplement $\Gamma_k^{\mathrm{Matter}}$ with a standard gauge-fixing term for the vector field $A_\mu^a$. Note that we are treating all vectors as Abelian gauge fields; this is justified since our analysis is restricted to the fixed-point regime and we do not explore fixed-point solutions with interacting gauge couplings.

The third term in \cref{eq:truncation_full} includes induced interaction. These interactions are unavoidably present at the fixed-point regime as they share the maximal global symmetries of $\Gamma_k^{\mathrm{Matter}}$. In this work, we consider the following contributions to our ansatz
\begin{align}
\Gamma_k^{\mathrm{Induced}} &= 
\int \mathrm d^4 x   \sqrt{g}\, \bigg(  \frac{Z_\psi^2 k^{-2}\lambda_+}{2} \Big( \big(\bar{\psi}_i \gamma^\mu \psi_i \big)^2 +  \big(i \bar{\psi}_i \gamma^\mu \gamma^5\psi_i \big)^2 \Big) \notag \\
&\qquad\qquad\qquad+ Z_\psi k^{-2} \sigma_\text{Ric}\,  R^{\mu\nu} \,\big( i \bar{\psi}_i \gamma_\mu D_\nu \psi_i - i D_\mu \bar{\psi}_i \gamma_\nu  \psi_i \big) 
+ Z_\psi k^{-2} \sigma_R\, R \, \big( i \bar{\psi}_i \gamma^\mu D_\mu \psi_i - i D_\mu \bar{\psi}_i \gamma^\mu  \psi_i \big)   \notag\\
&\qquad\qquad \qquad
+ Z_\phi k^{-2} \rho_\text{Ric}\,R^{\mu\nu}\, \partial_\mu\phi_I\,  \partial_\nu\phi_I +  Z_\phi k^{-2} \rho_\text{R}\,R\, \partial_\mu\phi_I\,  \partial^\mu\phi_I  \bigg),
\end{align}
where those terms with derivatives acting on fermions were anti-symmetrised to account for Osterwalder-Schrader positivity. Note that we are writing explicitly $k$-dependent terms to ensure that all couplings are dimensionless.

Finally, the third term in \cref{eq:truncation_full} includes operators that share the same symmetries of Yukawa interactions. We work with an approximation where only one of the scalar fields, namely $\phi_1$, interacts with fermions via a Yukawa term. Therefore, the underlying symmetry used to construct our ansatz for $\Gamma_k^{\mathrm{Yukawa}}$ is $\phi_1 \mapsto - \phi_1$, $\psi_i \mapsto \gamma_5 \psi_i$, and $\bar{\psi}_i \mapsto - \bar{\psi}_i \gamma_5$. Here, we consider the following contributions to our ansatz
\begin{align}
\Gamma_k^{\mathrm{Yukawa}} &=
Z_{\phi}^{1/2} Z_\psi \int \mathrm d^4 x   \sqrt{g}\, \bigg[\, i y \, \phi_1 \, \psib_i \psi_i +  i  k^{-2} y_\Box \, \phi_1\, \big( \psib_i \Box\psi_i  + \Box\psib_i \psi_i \big) \notag \\
&\qquad\qquad\qquad\qquad\, + i  \phi_1 \, \psib_i \psi_i \Big( k^{-2} y_R \, R  + k^{-4} y_{R^2} \, R^2 + k^{-4} y_{C^2} C_{\mu\nu\alpha\beta}^2 \Big)\notag  \\
&\qquad\qquad\qquad\qquad\, + i \phi_1 \, \psib_i \psi_i \left( \frac{1}{2} \, Z_\psi k^{-4} y_{\psib\psi}\, \big( i D_\mu \psib_j \gamma^\mu  \psi_j - i \psib_j \gamma^\mu  D_\mu\psi_j \big)   
+ \frac{1}{4} Z_A k^{-4} y_{A^2} \,F_{\mu\nu}^a F^{a\mu\nu} \right)\bigg] \,.
\end{align}
Since the terms that are present in $\Gamma_k^{\mathrm{Yukawa}}$ break the maximal global symmetries of the $\Gamma_k^{\mathrm{Matter}}$, they are not induced at the fixed-point regime. Therefore, we can focus on fixed-point solutions where the (generalised) Yukawa couplings $(y,\,y_\Box,\,y_R,\,y_{R^2},\,y_{C^2},\,y_{\psib\psi},\,y_{A^2})$ are all zero. However, these terms still play an important role in the discussion of the critical exponent for the Yukawa coupling as they contribute to NLO effects via off-diagonal terms of the stability matrix. 

Our ansatz for $\Gamma_k$ is organised such that all operators that contribute to $\Theta_y$ at NLO are included. This implies that our ansatz includes at most dimension-8 operators or operators with five fields. However, it is important to emphasise that the ansatz above defined does not include all possible operators up to dimension-8 that are compatible with the underlying symmetries of our system, namely those that do not contribute at NLO. Here we comment on the absence of some of those:
\begin{itemize}
\item We avoid higher-order kinetic terms as they lead to spurious poles in the propagators due to a finite order truncation. Instead, we account for higher-order corrections to the propagators in terms of momentum-dependent anomalous dimensions. In this way, we are partially accounting for the effects of resummed higher-order kinetic terms, including those with dimension higher than 8, without necessarily adding extra poles to the propagators. For example, the operators $R^2$ and $C_{\mu\nu\rho\sigma}^2$ contribute at NLO through the graviton propagator and this contribution is taken into account through $\eta_h(p^2)$. The vertex contribution of $R^2$ and $C_{\mu\nu\rho\sigma}^2$ is NNLO, and we neglect this.
\item Apart from the propagator contribution, we neglect higher-dimensional operators in $\Gamma_k^{\mathrm{Grav}}$, since these operators only contribute beyond NLO. Additionally, the effect of such operators would be to change the fixed-point values of the gravitation couplings $\gnewton$ and $\lambda$, and this is already taken into account in our analysis in the $\lambda$-$\gnewton$  plane that does not rely on a particular fixed-point value for the gravitational couplings.
\item We neglect the dimension-8 operators in the lower half of \cref{tab:induced} although they are contributing at NLO. These couplings tend to be small at the fixed-point regime and only contribute through the matter anomalous dimensions, which are typically small. Moreover, some of the induced dimension-8 operators, such as $(\partial_\mu \phi \partial^\mu \phi)^2$, may lead to spurious fixed-point collisions \cite{Eichhorn:2011pc, deBrito:2021pyi, Laporte:2021kyp, Knorr:2022ilz, deBrito:2023myf}, therefore, compromising our analysis.\\
We also neglect the induced operator $\sigma_{\nabla \!R}  \,  \nabla^\mu R \,\,  i \bar{\psi}_i \gamma_\mu \psi_i$, because our projection scheme to extract the flow of non-minimal couplings in the fermionic sector only allow us to disentangle two independent couplings as a result of the choice of kinematic configuration.
\item There are other dimension-6 and -8 operators that are compatible with the symmetries used to define $\Gamma_k^{\mathrm{Yukawa}}$, for example, there are two more channels of operators with structure similar to one associated with $y_\Box$, seven more channels of operators with structure similar to the one associated with $y_{\psib\psi}$ and two operators of the form $(\partial\phi)^2\phi_1 \psib\psi$. We explicitly tested the impact of these terms in our analysis, but their contribution vanishes at NLO due to momentum or Dirac structure. In consequence, their contribution is sub-leading and we neglect them. \\
Moreover, we could also add operators similar to the one associated with $y_\Box$, but with higher derivatives. Their contribution is partially included in the momentum-dependent anomalous dimensions $\eta_\phi(p^2)$ and $\eta_\psi(p^2)$; therefore, we neglect their explicit contributions.
\item Finally, we note that the operator $y_E E\, \phi_1 \psib_i \psi_i$, where $E = R^2 - 4 R_{\mu\nu}^2 + R_{\mu\nu\rho\sigma}^2$ is the Gau\ss-Bonnet density, is not topological. Nonetheless, the contribution from this operator only enters at NNLO. Looking at the $h^n\phi \bar\psi\psi$ vertex from this operator, the vertex vanishes identically when all graviton momenta sum to zero $\sum_{i=1}^n p_{h_i} = 0$ since this is precisely the kinematic limit in which the term falls back to a topological invariant. Looking at the flow of the Yukawa coupling, the NLO operators only contribute via the tadpole diagram. The $h^2\phi \bar\psi\psi$ vertex in this diagram has precisely $p_{h_1} = - p_{h_2}$ and therefore the $y_E$ contribution is vanishing at NLO. 
\end{itemize}

\subsection{Comment on derivation of beta functions}

In this work, we have analysed a system of (generalised) Yukawa couplings $(y,\,y_\Box,\,y_R,\,y_{R^2},\,y_{C^2},\,y_{\psib\psi},\,y_{A^2})$ under the influence of gravitational interactions. In this section of the supplementary material, we provide a concise discussion on the derivation of the beta functions used in our analysis.

We compute the beta functions from the flow of $n$-point functions derived from the Wetterich equation, where the number and type of field depends on the specific coupling we are focusing on. For example, the beta functions of $y$ and $y_\Box$ were extracted from the three-point Yukawa vertex, the beta function of $y_R$ was extract from the four-point Yukawa vertex with an additional graviton, the beta functions of $y_{R^2}$ and $y_{C^2}$ were extracted from the five-point Yukawa vertex with two additional gravitons, and $y_{\psib\psi}$ and $y_{A^2}$ were extracted from the five-point Yukawa vertex with two additional fermions and gauge fields, respectively. Schematically, the beta functions are given by
\begin{subequations}
\begin{align}
	\beta_y & = \left(\eta_\psi(0)+ \frac12 \eta_\phi(0)\right) y + \text{Flow}_{y} \,,\\
	%%%%
	\beta_{y_{\Box}} & = \left(2+\eta_\psi(0) + \frac12 \eta_\phi(0) \right) y_{\Box} + \frac{1}{4} \left( \eta'_\psi(0) + \eta'_\phi(0)\right) y  + \text{Flow}_{y_\Box} \,, \\[1ex]
	%%%%
	\beta_{y_R} & = \left(2+\eta_\psi(0) + \frac12 \eta_\phi(0)+ \frac12 \eta_{h}(0) \right) y_R + \frac{1}{2} \left(\eta'_\psi(0) + \eta'_{h}(0)\right) y   +\left(1- \frac{\beta_g}{2g}\right) y_R 
	+ \text{Flow}_{y_R}  \,, \\[1ex]
	%%%%
	\beta_{y_{R^2}} & = \left(4+\eta_\psi(0) + \frac12 \eta_\phi(0)+ \eta_{h}(0)\right) y_{R^2} +  \frac{1}{6}\eta'_{h}(0)\, y_R + \frac{1}{12} \,\eta''_{h}(0) \,y + \left(2- \frac{\beta_g}{g}\right) y_{R^2}
	+ \text{Flow}_{y_{R^2}}  \,, \\[1ex]
	%%%%
	\beta_{y_{C^2}} &= \left(4+\eta_\psi(0) + \frac12 \eta_\phi(0) + \eta_{h}(0)\right) y_{C^2} - \frac{1}{2} \eta'_{h}(0) \,y_R - \frac{1}{4} \eta''_{h}(0) \,y  +\left(2- \frac{\beta_g}{g}\right) y_{C^2}+
	\text{Flow}_{y_{C^2}}  \,, \\[1ex]  
	%%%%
	\beta_{y_{A^2}} &= \left(4+\eta_\psi(0) + \frac12 \eta_\phi(0) + \eta_{A}(0) \right) y_{A^2}  + \text{Flow}_{y_{A^2}}\,, \\[1ex]
	%%%%
	\beta_{y_{\psib\psi}} &= \left(4+2\,\eta_\psi(0) + \frac12 \eta_\phi(0) \right) y_{\psib\psi}  + \text{Flow}_{y_{\psib\psi}} \,,
\end{align}
\end{subequations}
where we denote $\text{Flow}_{X}$ as the sum of diagrams obtained by taking functional derivatives of the right-hand side of the Wetterich equation, and then projected onto the coupling $X$.

In general, the beta functions depend on anomalous dimensions of the fields.  In this work, we compute the momentum-dependent $(\eta_h(p^2),\eta_\phi(p^2),\eta_\psi(p^2),\eta_A(p^2))$ anomalous dimensions from the flow of two-point functions and employing the so-called bilocal approximation. See, for instance, \cite{Christiansen:2012rx, Christiansen:2014raa, Meibohm:2015twa} for discussions on the derivation of anomalous dimension within the bilocal approximation. Note that in our bilocal approximation of the anomalous dimension, $\eta''_\Phi(p)=0$.

The influence of the gravitational interactions on the flow of the Yukawa couplings is described through its dependence on the Newton coupling and cosmological constant. Therefore, we also computed the flow of $\gnewton$ and $\lambda$. The flow of $\gnewton$ was extracted from the graviton three-point function, as discussed in \cite{Christiansen:2015rva, Denz:2016qks, Pawlowski:2020qer}. Moreover, we use the graviton two-point function to extract the flow of $\lambda$, but setting $\lambda = 0$ in all graviton $n$-point functions with $n\geq3$. This approximation is motivated by previous studies showing that the fixed-point value of $\lambda$ obtained from a flow extracted from graviton $n$-point functions, with $n\geq3$, is significantly smaller than the fixed-point value of $\lambda$ obtained from a flow extracted from the graviton two-point function \cite{Christiansen:2015rva, Meibohm:2015twa, Denz:2016qks}.

In addition to the system of Yukawa couplings, our analysis also involves the beta functions of the induced couplings $(\lambda_+,\rho_\text{Ric},\rho_R,\sigma_\text{Ric},\sigma_R)$. The beta function for the four-fermion coupling $\lambda_+$ has been previously computed in \cite{Eichhorn:2011pc, Meibohm:2016mkp, deBrito:2020dta, deBrito:2023kow}. In this work, we extend the computation $\beta_{\lambda_+}$ to include contributions of the non-minimal couplings in the fermion sector. For the non-minimal couplings, we use finite momentum projection, which better captures the momentum-dependent flow of the vertices $h\psi\psib$ and $h\phi\phi$ in comparison to the derivative expansion. See \cite{Eichhorn:2018nda} for a discussion on the momentum dependence of the three-point vertex $h\psi\psib$. Schematically, the beta functions are given by
\begin{subequations}
\begin{align}
	\beta_{\rho_\text{Ric}} &= 2\rho_\text{Ric} - \frac{1}{2}\left[ \frac{1}{2} \big( \eta_h(k^2) - \eta_h(0)\big) + \eta_\phi(k^2) - \eta_\phi(0)\right]  \left( 2 - 3 \rho_\text{Ric} + \rho_\text{Ric}^2 \right) \notag\\
	&\qquad+ \big(2-\rho_\text{Ric}\big)\, \text{Flow}_{\rho_\text{Ric}}(k^2) - \big(4-4\rho_\text{Ric}\big)\, \text{Flow}_{\rho_\text{Ric}}(k^2/2) \,,\displaybreak[0]\\[2ex]
	%%%%%%%%%%%%%%%%%%%%%%%%%%%%%%%%%%%%%%%%%%%%%%%%%%%%%%%%%%%%%%%%%%%%%%%%%%%%%%%%
	\beta_{\rho_R} &= 2\rho_R + \frac{1}{2}\left[ \frac{1}{2} \big( \eta_h(k^2) - \eta_h(0)\big) + \eta_\phi(k^2) - \eta_\phi(0)\right]  \big( 1 +3 \rho_R + 2 \rho_R^2 \big) \notag \\
	&\qquad+ \big(1+\rho_R\big)\, \text{Flow}_{\rho_R}(k^2) - \big(2+4\rho_R\big)\, \text{Flow}_{\rho_R}(k^2/2) \,, \displaybreak[0]\\[2ex]
	%%%%%%%%%%%%%%%%%%%%%%%%%%%%%%%%%%%%%%%%%%%%%%%%%%%%%%%%%%%%%%%%%%%%%%%%%%%%%%%%
	\beta_{\sigma_\text{Ric}} &=  2\sigma_\text{Ric} - \frac{1}{2} \left[ \frac{1}{2} \big( \eta_h(k^2) - \eta_h(0)\big) + \eta_\psi(k^2) - \eta_\psi(0)\right] \big( 1 - 3\sigma_\text{Ric} + 2 \sigma_\text{Ric}^2  \big) \notag \\
	&\qquad- \big(1-\sigma_\text{Ric}\big)\, \text{Flow}_{\sigma_\text{Ric}}(k^2) + \big(2-4\sigma_\text{Ric}\big)\, \text{Flow}_{\sigma_\text{Ric}}(k^2/2) \,, \displaybreak[0]\\[2ex]
	%%%%%%%%%%%%%%%%%%%%%%%%%%%%%%%%%%%%%%%%%%%%%%%%%%%%%%%%%%%%%%%%%%%%%%%%%%%%%%%%
	\beta_{\sigma_R} & = 2\sigma_R +  \frac{1}{60} \left[ \frac{1}{2} \big( \eta_h(k^2) - \eta_h(0)\big) + \eta_\psi(k^2) - \eta_\psi(0)\right] 
	\big( 25  +15\sigma_\text{Ric} + 90 \sigma_R + 2 \sigma_\text{Ric}^2 + 72 \sigma_R^2 +24 \sigma_\text{Ric} \sigma_R \big)  \notag \\
	&\qquad+ \big(5 + \sigma_\text{Ric} + 6 \sigma_R\big)\, \text{Flow}_{\sigma_R}(k^2) - 2\big(5 +2\sigma_\text{Ric} + 12 \sigma_R\big)\, \text{Flow}_{\sigma_R}(k^2/2) 
	+\frac{1}{6} \big( 2 \sigma_\text{Ric} - \beta_{\sigma_\text{Ric}} \big)\,,
\end{align}
\end{subequations}
where $\text{Flow}_X(p^2)$ denotes a momentum-dependent flow contracted with an appropriate tensor structure to project it onto the coupling $X$. See \cite{Eichhorn:2018nda} for details on how to define this kind of projection. A subset of these beta functions, namely $\beta_{\rho_\text{Ric}}$ and $\beta_{\sigma_\text{Ric}}$, was partially computed in \cite{Eichhorn:2017sok, Eichhorn:2018nda}. The present work is the first computation of this set of beta functions, including their combined effects.

The explicit computations of the beta functions discussed in this section relied on self-written \textit{Mathematica} codes \cite{AScodebase} integrated with the packages \textit{xAct} \cite{Brizuela:2008ra, Martin-Garcia:2007bqa, Martin-Garcia:2008yei, Martin-Garcia:2008ysv}, \textit{FormTracer} \cite{Cyrol:2016zqb} and \textit{DoFun} \cite{Huber:2011qr, Huber:2019dkb}. Part of the computations involving large tensor structures was also done with \textit{FORM} \cite{Vermaseren:2000nd, Kuipers:2012rf}.

%%%%%%%%%%%%%%%%%%%%%%%%%%%%%%%%%%%%%%%%%%%%%%%%%%%%%%%%%%%%%%%%%%%%%%%%%%%%%%%%
%%%%%%%%%%%%%%%%%%%%%%%%%%%%%%%%%%%%%%%%%%%%%%%%%%%%%%%%%%%%%%%%%%%%%%%%%%%%%%%%

Finally, let us comment on the choice of regulator. When computing the beta function for the generalised Yukawa couplings $y_{R^2}$, $y_{C^2}$ and $y_{A^2}$, we use a projection scheme that requires taking four momentum derivatives of their respective flows. In this case, the standard Litim regulator \cite{Litim:2000ci} is not a suitable choice since it leads to unphysical divergences due to the non-cancellation of terms involving four derivatives of Dirac deltas. 

A convenient choice, that was employed in \cite{Balog:2019rrg} to handle a similar problem, is
\begin{align}
\label{eq:Rk_boson}
R^{(n)}_k (p) = k^2 \,(1 - p^2/k^2)^n \,\theta(1 - p^2/k^2) \,,
\end{align}
to regularise bosonic propagators. Note that we recover the standard Litim regulator for $n=1$.  To deal with fermionic propagators, we use the regulator
\begin{align}
\label{eq:Rk_fermion}
R^{(n)}_{k,\text{F}} (p) = \slashed{p} 
\Bigg[ -1 + \bigg( \frac{k^2}{p^2} \left( 1 - \frac{p^2}{k^2} \right)^{\!n} + 1 \bigg)^{1/2} \,\Bigg] \,\theta(1 - p^2/k^2) \,,
\end{align}
which is defined such that 
\begin{align}
\text{``Regularized fermionic dispersion''} = \sqrt{\text{``Regularized bosonic dispersion''}} \,.
\end{align}
The main idea of using this class of regulator is that, by increasing the value of $n$, we can make the regulator sufficiently ``smooth'' such that we can eliminate the unphysical divergences due to delta peaks. In our computation, we used \cref{eq:Rk_boson,eq:Rk_fermion} with $n=2$, which is sufficient for our purposes.

%%%%%%%%%%%%%%%%%%%%%%%%%%%%%%%%%%%%%%%%%%%%%%%%%%%%%%%%%%%%%%%%%%%%%%%%%%%%%%%%
%%%%%%%%%%%%%%%%%%%%%%%%%%%%%%%%%%%%%%%%%%%%%%%%%%%%%%%%%%%%%%%%%%%%%%%%%%%%%%%%
%%%%%%%%%%%%%%%%%%%%%%%%%%%%%%%%%%%%%%%%%%%%%%%%%%%%%%%%%%%%%%%%%%%%%%%%%%%%%%%%
%%%%%%%%%%%%%%%%%%%%%%%%%%%%%%%%%%%%%%%%%%%%%%%%%%%%%%%%%%%%%%%%%%%%%%%%%%%%%%%%

\subsection{Fixed point and anomalous dimensions}
\label{app:critexps}
We list all fixed-point values, anomalous dimensions and critical exponents in this supplement.

\subsubsection{Minimal matter}
For $N_{s}=N_\text{f} = 1$ and $N_\text{v}=0$, we find a fully interacting fixed point at the values
\begin{align}
(\gnewton,\, \lambda,\, \rho_\text{Ric},\, \rho_\text{R},\, \sigma_\text{Ric},\, \sigma _\text{R},\, \lambda_+)^*
= 
(0.67,\, 0.19,\, -0.12,\, 0.50,\, -0.030,\, 0.030,\, -1.0)\,,
\end{align}
with the critical exponents
\begin{align}
\theta_i	 = (1.3 \pm 5.4\, i,\, - 1.7,\, -1.8,\, -2.4,\, -4.8 \pm 0.066\, i)\,.
\end{align}
The two relevant directions belong to $\gnewton$ and $\lambda$ while all higher-order operators remain irrelevant. The anomalous dimensions are given by
\begin{align}
( \eta_h (k^2),\, \eta_h(0),\, \eta_\phi (k^2),\, \eta_\phi(0),\, \eta_\psi (k^2),\, \eta_\psi(0))^*
= 
(0.098,\, 1.3,\, 0.63,\, -0.23,\, 0.062,\, -0.45)\,.
\end{align}
All Yukawa couplings have an asymptotically free fixed point
\begin{align}
y^*=y_R^*=y_\Box^*=y_{R^2}^*  =y_{C^2}^* = y_{\bar\psi\psi}^* = 0 \,.
\end{align}
The critical exponents in the Yukawa sector are given by
\begin{align} 
\theta_{y_i} = (3.1,\, -2.6 \pm 0.33\, i,\, -2.7 \pm 7.7\, i,\, -3.8)\,.
\end{align}
The eigenvector of the relevant direction is clearly aligned with the standard Yukawa coupling.

\subsubsection{SM matter content}
For SM matter content, we find a fully interacting fixed point at the values
\begin{align}
(\gnewton,\, \lambda,\, \rho_\text{Ric},\, \rho_\text{R},\, \sigma_\text{Ric},\, \sigma _\text{R},\, \lambda_+)^*
= 
(0.17,\, 0.23,\, -0.18,\, 0.37,\, -0.039,\, 0.039,\, -0.11)\,,
\end{align}
with the critical exponents
\begin{align}
\theta_i	 = (1.7 \pm 5.7\, i ,\, - 1.7 \pm  0.44\, i,\,  - 1.8,\, -1.9,\, - 2.0)\,.
\end{align}
The two relevant directions belong to $\gnewton$ and $\lambda$ while all higher-order operators remain irrelevant.  The anomalous dimensions are given by
\begin{align}
&( \eta_h (k^2),\, \eta_h(0),\, \eta_\phi (k^2),\, \eta_\phi(0),\, \eta_\psi (k^2),\, \eta_\psi(0),\, \eta_A (k^2),\, \eta_A(0))^* \notag \\[1ex]
&\hspace{3cm}= 
(0.064,\, 0.73,\, 0.13,\, -0.050,\, 0.022,\, -0.17,\, -0.052,\, 0.019)\,.
\end{align}
All Yukawa couplings have an asymptotically free fixed point
\begin{align} \label{eq:Yukawa-ordering}
y^*=y_R^*=y_\Box^*=y_{R^2}^*  =y_{C^2}^* = y_{\bar\psi\psi}^* = y_{A^2}^* = 0 \,.
\end{align}
The critical exponents in the Yukawa sector are given by
\begin{align}
\theta_{y_i} = (2.2,\, -2.2,\, -3.4,\, -3.7, -4.0 \pm 6.4\, i,\, -4.2 )\,.
\end{align}
The eigenvector of the relevant direction is clearly aligned with the standard Yukawa coupling.

\subsection{Estimate of systematic uncertainties due to higher-order operators}
\label{app:simulations}
In the main part, we have presented error estimates for the critical exponents of the Yukawa couplings $y_i$. These error estimates are informed by the structure of the beta-functions at higher orders. The operators investigated so far are at most five-field operators, since an NLO operator needs to directly contribute to $\beta_y$. There are no additional five-field operators that contribute to $\beta_y$, apart from adding explicit derivatives to existing operators. Hence, we now estimate the impact of six-field NNLO operators. These include, for example, $R^3\phi \bar\psi\psi$.
Compared to the NLO operators, there are two crucial differences:
\begin{itemize}
\item[i)] There is no direct contribution from an NNLO operator to $\beta_y$ and $\beta_{y_\Box}$. These operators only contribute to vertices with at least six fields, and the highest order vertex in the Yukawa beta function contains five fields.
\item[ii)] The operators have canonical mass dimension 10 (or higher) and therefore the diagonal element contains the factor $6 + \# G_\text{N}$.
\end{itemize}
Further, we note that the $\gnewton^2$ contribution to $\Theta_y$ from non-induced operators typically follows the form described in \cref{eq:stab-mat-schematic}, where both off-diagonal contributions are $\sim \gnewton$. This however is not true for the couplings
$y_{\bar\psi\psi}$ and $y_{A^2}$, where the $M_{1j}$ entry is $\gnewton$ independent, while the $M_{j1}$ entry starts at $\gnewton^2$. However, we can always rescale $y_{\bar\psi\psi}\to \gnewton\, y_{\bar\psi\psi}$ and $y_{A^2}\to \gnewton\, y_{A^2}$, which by virtue of mere rescalings leaves the critical exponents unchanged. This rescaling has the advantage that now all off-diagonal entries of the stability matrix $M$ start at linear order in $\gnewton$.

Keeping this possibility of rescaling in mind, and focusing on the full set of NLO operators plus a single additional NNLO six-field operator, the stability matrix takes the form
\begin{align}
M
=
\begin{pmatrix}
	&	& & & & & & 0\\
	&	& & & & & &\#  \gnewton\\
	&	& & & & & &0\\
	&	& & & & & &\#  \gnewton\\
	\multicolumn{7}{c}{\smash{\raisebox{1\normalbaselineskip}{\large $M_\text{NLO}$}}}	&\#  \gnewton	\\
	&	& & & & & &\#  \gnewton\\
	&	& & & & & &\#  \gnewton\\
	\#  \gnewton  &\#  \gnewton  &\#  \gnewton  &\#  \gnewton  &\#  \gnewton &\#  \gnewton &\#  \gnewton  & 6 + \#  \gnewton 
\end{pmatrix}\,,
\end{align}
where $M_\text{NLO}$ is the $7\times7$ stability matrix of all NLO operators, and where the zero-entries in the last column reflect the absence of direct contributions to $\beta_y$ and $\beta_{y_\Box}$.

Instead of computing the entries $\#$ in the above stability matrix, we perform a Monte-Carlo simulation to estimate the effect of a \textit{typical} higher-order operator. For this, we replace each entry $\#$ by a random number, i.e., $\#\to \mathrm{rand[-range, range]}$. In particular, we choose a uniform distribution, and still need to determine the range of the random numbers. First, we assume that the higher-order operator has entries of similar order as the ones included in $M_\text{NLO}$. While this is an assumption, this pattern is true for $M_\text{NLO}$ itself, which does contain four- and five-field operators of different mass dimension, and no clear tendency of the size of entries is visible. Second, we recall that entries of the stability matrix are not physical, in the sense that a linear transformation $g_i\to a_i \, g_i$ leaves the critical exponents unchanged, but changes the stability matrix. Furthermore, for couplings of the same kind, in particular for $\yukRsq$ and $\yukCsq$, we allow for linear combinations of both operators. To fix this arbitrariness on the level of our chosen basis and parametrisation, we homogenise the entries of $M_\text{NLO}$: we adjust the parameters $a_i$ such that the standard deviation of all non-vanishing entries ($\sim \gnewton$) of $M_\text{NLO}$ is minimised. This ensures that all NLO operators contribute as equally as possible, and are not over-represented by a particular choice of ours.  From this homogenised $M_\text{NLO}$, we choose the largest $\sim \gnewton$ coefficient to determine the range of our random numbers.

To illustrate this procedure, let us focus on the case of SM matter content, and fixing $\lambda=\lambda_*=0.23$. The stability matrix $M_\text{NLO}$ to linear order in $\gnewton$ with the same ordering of couplings as in \cref{eq:Yukawa-ordering} reads
\begin{align}
M_\text{NLO}=	\begin{pmatrix}
	1.3 \, \gnewton	& 1.1 \,\gnewton	& 0.54\, \gnewton & 1.3\, \gnewton & -2.2 \,\gnewton &-0.14 \,\gnewton & -0.048 \,\gnewton \\
	-23 \, \gnewton & 2- 0.73 \, \gnewton & -5.5\, \gnewton & 0.46\, \gnewton & 8.5 \, \gnewton & -0.014 \, \gnewton & -0.0093 \, \gnewton \\
	3.0 \, \gnewton & 1.6 \, \gnewton & 2+1.8\, \gnewton &0 &0 &0 &0 \\
	95 \, \gnewton & 36 \, \gnewton & 25 \, \gnewton & 4+ 2.4 \, \gnewton & -43 \, \gnewton & 0 & 0\\
	31 \, \gnewton & 13 \, \gnewton & 5.7 \, \gnewton & 34 \, \gnewton & 4-5.9\, \gnewton & 0& 0\\
	-1100 \, \gnewton & -680 \, \gnewton & -370 \, \gnewton & 630 \, \gnewton & 410 \, \gnewton & 4-0.39\, \gnewton & 0\\
	-150 \, \gnewton & -46 \, \gnewton & -23 \, \gnewton & 5.5 \, \gnewton & 18 \, \gnewton & 0 & 4- 1.4 \, \gnewton 
\end{pmatrix}+ \mathcal{O}(\gnewton^2)\,,
\end{align}
which, after rebalancing as described above, transforms into

\begin{align}
\label{eq: NLOrebal}
\tilde{M}_\text{NLO}=	\begin{pmatrix}
	1.3 \, \gnewton	& 2.3\,\gnewton	& -1.5\, \gnewton &-4.0\, \gnewton & -8.7 \,\gnewton &-21 \,\gnewton & -0.62 \,\gnewton \\
	-11 \, \gnewton & 2- 0.75 \, \gnewton & 6.9\, \gnewton & -14\, \gnewton & 5.6 \, \gnewton & -1.0 \, \gnewton & -0.056 \, \gnewton \\
	-1.1 \, \gnewton & -1.3 \, \gnewton & 2+1.8\, \gnewton &0 &0 &0 &0 \\
	-16 \, \gnewton & -13 \, \gnewton & 10 \, \gnewton & 4+ 3.0 \, \gnewton & 39 \, \gnewton & 0 & 0\\
	-14 \, \gnewton & -11 \, \gnewton & -11 \, \gnewton & -37 \, \gnewton & 4-6.4 \, \gnewton & 0& 0\\
	-7.4 \, \gnewton & -9.5 \, \gnewton & -6.5 \, \gnewton & -20 \, \gnewton & -12 \, \gnewton & 4-0.38\, \gnewton & 0\\
	-12 \, \gnewton & -7.5 \, \gnewton &4.8 \, \gnewton &- 5.9 \, \gnewton & 0.62 \, \gnewton & 0 & 4- 1.4 \, \gnewton 
\end{pmatrix}+ \mathcal{O}(\gnewton^2)\,.
\end{align}
One can see that the coefficients linear in $\gnewton$ are more homogeneous, and one can check explicitly that the critical exponents remain unchanged. For the stability matrix given in \cref{eq: NLOrebal}, we therefore set $\mathrm{range}=37$ for the random numbers.

This procedure ensures that the impact of a typical higher-order operator can be estimated. For each simulation, we average over $10^5$  random stability matrices.
We perform two sets of analysis:
\begin{itemize}
\item[1)] We estimate the error on $\Theta_y$ as a function of $\gnewton$ at fixed cosmological constant, $\lambda=\lambda_*$. For a given value of $\gnewton$, we evaluate all critical exponents, sort them according to their real part, and compare them with the (sorted) critical exponents of the NLO system. The upper error is given by the standard-deviation of all $\Theta_{i,\, \mathrm{sim}}>\Theta_{i,\, \mathrm{NLO}}$, and the lower error by the standard-deviation of all $\Theta_{i,\, \mathrm{sim}}<\Theta_{i,\, \mathrm{NLO}}$. This procedure results in \cref{fig:ThetasNLOvsNNLO}, and  for $\gnewton=\gnewton^*$ in the error on $\Theta_y$, see \cref{eq:theta-Yuk-res}.
\item[2)] We estimate the uncertainty of the line separating the red and green regions in \cref{fig:G-Lambda-plot}. For a given value of $\lambda$, we increase $\gnewton$, and count how many of the simulations give $\Theta_y>0$. The lower error indicates when $16\%$ of the simulations lead to a relevant Yukawa coupling, and the upper error indicates when $84\%$ of the simulations lead to a relevant Yukawa coupling, corresponding to a $1\sigma$ region. 
\end{itemize}
\end{document}